\documentclass[opre]{informs3nothing}  

\DoubleSpacedXI 

\usepackage{latexsym, bbm, enumerate, amssymb, amsmath, libertine,tikz,multirow}
\usepackage{gensymb}
\usepackage{booktabs}
\usepackage[utf8]{inputenc}
\usepackage[T1]{fontenc}
\usepackage{csquotes}
\usepackage{booktabs}

\usepackage{natbib}
\bibpunct[, ]{(}{)}{,}{a}{}{,}%
\def\bibfont{\small}%
\graphicspath{{./figures/}}
\newcommand{\vx}{\mathbf{x}}
\newcommand{\vz}{\mathbf{z}}
\renewcommand{\Pr}{\mathbb{P}}

\TheoremsNumberedThrough     
\ECRepeatTheorems

\EquationsNumberedThrough    

\newcommand{\paranth}[1]{\left(#1\right)}

\newcommand{\curly}[1]{\left\{#1\right\}}

\TITLE{Detecting Bots and Assessing Their Impact in \\Social Networks}
\RUNTITLE{Detecting Bots and Assessing Their Impact}

\RUNAUTHOR{Guenon des Mesnards, et. al.}
\ARTICLEAUTHORS{%
	\AUTHOR{Nicolas Guenon des Mesnards}
	\AFF{%
Operations Research Center, Massachusetts Institute of Technology, 
77 Massachusetts Ave.,
Cambridge, MA~  02139,  
		\EMAIL{ngdm@mit.edu}
	}
	\AUTHOR{David Scott Hunter}
\AFF{%
	Operations Research Center, Massachusetts Institute of Technology, 
	77 Massachusetts Ave.,
	Cambridge, MA~  02139, 
	\EMAIL{dshunter@mit.edu}
}
\AUTHOR{Zakaria el Hjouji}
	\AFF{%
	Massachusetts Institute of Technology, 
	77 Massachusetts Ave.,
	Cambridge, MA~  02139, 
	\EMAIL{zakariae@mit.edu}
}

	\AUTHOR{Tauhid Zaman}
	\AFF{%
		Yale School of Management, Yale University,  
		165 Whitney Ave, New Haven, CT 06511  
		\EMAIL{tauhid.zaman@yale.edu}
	}
} 

\ABSTRACT{Online social networks are often subject to influence campaigns by malicious actors through the use of automated accounts known as bots.   We consider the problem of detecting bots in online social networks and assessing their impact on the opinions of individuals.  We begin by analyzing the behavior of bots in social networks and identify that they exhibit heterophily, meaning they interact with humans more than other bots.  We use this property to develop a detection algorithm based on the Ising model from statistical physics.  The bots are identified by solving a minimum cut problem.  We show that this Ising model algorithm can identify bots with higher accuracy while utilizing much less data than other state of the art methods.  
   
   We then develop a   function we call \emph{generalized harmonic influence centrality} to estimate the impact bots have on the opinions of users in social networks.  This function is based on a generalized opinion dynamics model and captures how the activity level and network connectivity of the bots shift equilibrium opinions.  To apply generalized harmonic influence centrality to real social networks, we develop a deep neural network to measure the opinions of users based on their social network posts.  Using this neural network, we then calculate the generalized harmonic influence centrality of bots in multiple real social networks.  For some networks we find that a limited number of bots can cause non-trivial shifts in the population opinions.   In other networks, we find that the bots have little impact.  Overall we find that generalized harmonic influence centrality is a useful operational tool to measure the impact of bots in social networks.
}

\begin{document}
\maketitle
\section{Introduction}

Social networks face the challenge posed by automated bots which
create spam and result in a degraded user experience.  However,
recently these bots have become a serious threat to democracies.
There have been multiple reports alleging that foreign actors attempted to penetrate U.S. social networks in order to manipulate elections \citep{ref:russianbots,shane2017fake,guilbeault2016twitter,byrnes2016bot,ferrara2017disinformation}.  The perpetrators used bots to share  politically polarizing content, much of it fake news, in order to amplify it and extend its reach, or directly interacted  with humans to promote their agenda.    While no one knows exactly how many people were impacted by these influence campaigns, it has still become a concern for the governments of the U.S. and many other nations \citep{ref:russianbots_govtresponse,ref:russianbots_feinstein}.    

Social network counter-measures are needed to combat these coordinated influence campaigns.  Conventional methods of bot detection may not be sufficient because they can be fooled by modifying certain elements of bot behavior, such as the frequency of posting or sharing content.  However, because many of these bots are coordinated, they may exhibit joint behaviors which are difficult to mask and which allow for more accurate and robust detection.  These behaviors may not be observable by looking at accounts in isolation.  Therefore,   conventional algorithms which focus on individual detection may not find these bots.  What is needed  is an algorithm that can simultaneously detect multiple bots.


The potential threat to election security from social networks has become a concern for the U.S. government.  Facebook has identified several pages and accounts tied to foreign actors \citep{ref:russianbots} and Twitter suspended over 70 million bot accounts \citep{ref:twitter_70mbots}. One important question remains unanswered: what was the  impact  of these influence campaigns. More specifically, how do we quantify the impact of bots on the opinions of users in a social network? If we could do this, we would be able to assess the potential threat of a bot based influence campaign.

\subsection{Information Operations}\label{sec:info_ops}
\begin{figure}[!h]
	\centering
	\includegraphics[scale=0.6]{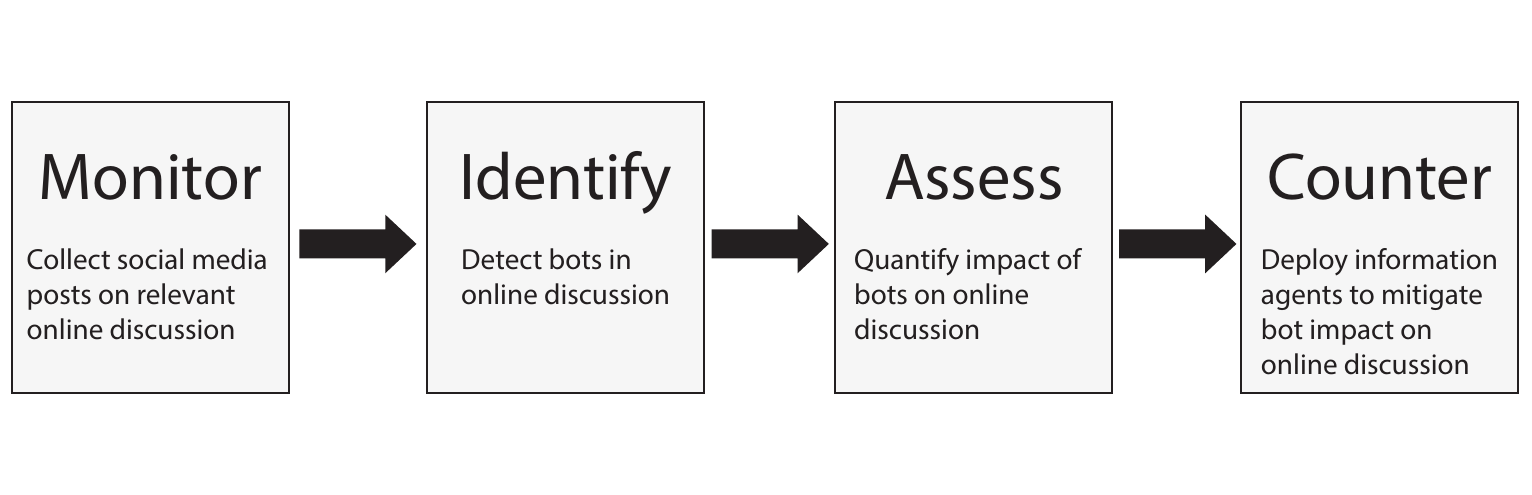} \\
	\caption{Illustration of the information operations workflow in online social networks.  Each box contains a different capability and  details on how it   applies to adversarial influence campaigns run by bots.  } \label{fig:info_ops}
\end{figure}
The bot detection and impact assessment questions fall under the umbrella of information operations.  This area focuses on tactics, techniques, and procedures used to achieve strategic objectives in the information environment \citep{dod2012joint}.  While the definition of the information environment is broad, in this work we focus on online social networks which are an increasingly important component of it.  Strategic objectives in online social networks are typically centered around detecting and mitigating influence campaigns run by adversaries.  With respect to these types of campaigns, information operations can be broken down into four main capabilities:  monitor, identify, assess, and counter.  The natural workflow of these capabilities that would be used in the field by practitioners is illustrated  in Figure \ref{fig:info_ops}.  We now discuss these capabilities and where our work fits in this information operations workflow.

	The first capability is monitoring, which refers to developing situational awareness by observing social network data.    In our work we focus on the online discussion surrounding different topics which we suspect to be targets for influence campaigns.  We collect social network posts about the topics and the network connections between the individuals posting.  Other monitoring capabilities may focus not only on topics, but also individuals in certain geographic regions.  In this case one would need the ability to geo-locate individuals.  There are many different approaches to geo-location in social networks  \citep{marks2017building, han2014text, jurgens2015geolocation}, but  we do not consider this here.

While monitoring posts about a topic, one would like to identify malicious actors in a social network who are engaged in the influence campaign.  These actors could be online extremists inciting violence \citep{klausen2018finding} or automated bots running influence campaigns \citep{davis2016botornot}.   In this work our interest is in identifying automated bots who are amplifying different messages.  We develop our own bot identification algorithm which allows us to identify coordinated groups of bots.  

Once we have identified the malicious actors in the network, we would like to know if their efforts are succeeding in influencing or shifting opinions.  This can be done with assessment capabilities which provide a quantitative  measurement of the impact of the malicious actors, who are bots in this work.  Impact can be defined in many ways.  Studies have looked at the volume of content produced by bots and their social network reach  during	the 2016 election \citep{bessi2016social}.  However, this approach
	does not indicate the effectiveness of the bots in shifting opinions. An approach has been proposed by \cite{aral2019protecting} to asses how much social media influence campaigns are affecting elections.  The authors' approach requires studying vast amounts of social media and voting data, much of which may not be easily accessible.

{Another approach to assess the impact of bots is to utilize the structure of the social network in which they operate. In this approach, one develops a function that calculates some sort of score for the bots in the network.  These functions are known as network centralities and they provide a simple way to measure the impact of bots, or more generally any nodes in a network.  The challenge is to find a centrality function whose value is closely related to the impact of the bots on the opinions in the network. The strength of the network centrality functions is that they take into account all the complex interactions in the network and they can be evaluated using publicly available data.  Our work here develops an assessment capability based on a particular network centrality function to assess the impact of bots.

The final capability in the information operations workflow is countering, which we take to mean  mitigating adversarial influence campaigns.  While we do not develop any countering capabilities in this work, it is an important component of the information operations workflow and the only component that requires active intervention in a social network.  One simple approach to countering malicious actors is to remove them from the social network.  This is what is done by large social media platforms such as Twitter or Facebook \citep{ref:twitter_70mbots, ref:russianbots}.  However, if one does not have such power over a social network, another approach is to deploy information agents into the network with the goal of undoing the influence caused by the malicious agents.  This is referred to as the maximizing influence problem and its different variations have been studied  by many authors \citep{kempe2003maximizing, kempe2005influential, vassio2014message,hunter2018opinion}.  The goal is to maximize some influence objective function by having the agent target individuals in the network with its messages.  Much of the literature focuses on algorithmic questions surrounding how to identify the targets.  However, \cite{hunter2018opinion} presented a practical implementation of a targeting algorithm on real social networks.

\subsection{Our Contributions}  
In this paper we present a method to identify bots in a social network and assess the impact they have on the opinions of users of the social network.  We begin by studying the behavior of bot accounts and identify key behaviors they exhibit.  We use this as the basis for a bot detection algorithm which models the bot interactions using the Ising model from statistical physics.  We show that bot detection can be reduced to solving a minimum cut problem.  Tests on real social networks show that our Ising model algorithm has a higher accuracy than other state of the art methods.

Having identified the bots, our next goal is to assess the impact they have on the opinions of social network users.  To do this, we model the interactions in a social network using a generalized opinion dynamics model which has been proven to reach an equilibrium determined by the network structure and the location of so-called \emph{stubborn} users whose opinions do not change.  Once a set of bots is identified,  we calculate the equilibrium opinions with and without the bots present.  The shift in these equilibrium opinions caused by the bots is how we quantify their impact.  The value of this shift is similar to a centrality function known as \emph{harmonic influence centrality} \citep{vassio2014message}. We modify this function so it can assess the joint impact of multiple nodes.  We call our function \emph{generalized harmonic influence centrality}.  We show how to operationalize this function so it can be applied to real social networks.  This involves the development of a deep neural network capable of measuring the opinions of social media posts.

We identify bots  using the Ising model algorithm in multiple real social networks with tens of thousands of users discussing geo-political issues.   Using generalized harmonic influence centrality, we find that the impact of bots varies across the networks and depends on factors such as the activity level of the bots, who they connect with, and the overall network structure.

This paper is outlined as follows.  We begin with a literature review in Section \ref{sec:lit_review}.  We then study the behavior of bots in Section \ref{sec:botsbehavior}.  This leads to the development of the Ising model algorithm for bot detection in Section \ref{sec:ising} and its performance evaluation in Section \ref{sec:results_ising}.  We present the opinion dynamics model and  generalized harmonic influence centrality in Section \ref{sec:assess}.  We apply  generalized harmonic influence centrality to real social networks to show the impact of bots in Section \ref{sec:assessment_results}.  We conclude in Section \ref{sec:conclusion}.


\section{Literature Review}\label{sec:lit_review}
\subsection{Bot Detection in Social Networks}
A detailed study of bots in the 2016 U.S. presidential election was
conducted by \cite{bessi2016social}.  The authors found a large fraction
of the election discussion came from bots that were connected to many users.  Similar conclusions were reached for bots
deployed in the run up to the Brexit vote \citep{bastos2019brexit}
and French elections \citep{ferrara2017disinformation}.  A comprehensive
survey of social bots is provided in \citep{ferrara2016rise}.  Social bots are designed to interact with other users \citep{hwang2012socialbots, messias2013you, boshmaf2013design} and post human-like content \citep{freitas2015reverse}. Not only do bots \emph{look} more human, but turn out to \emph{be} half-human. \cite{chu2012detecting} mention the concept of \emph{cyborgs}, where a real person manages dozens of otherwise automated accounts. Such hybrid accounts  make the detection task extremely challenging \citep{zangerle2014sorry, bbmg17}.

Bot detectors have become more sophisticated, from the earliest instances \citep{yardi2009detecting} to the state of the art \citep{davis2016botornot} currently used in many applications today \citep{ferrara2017disinformation, monsted2017evidence, vosoughi2018spread, badawy2018analyzing}. In \cite{ferrara2016rise} the authors present a taxonomy of bot detectors, from crowd-sourcing  \citep{wang2012social, stein2011facebook, elovici2014ethical} and honeypot traps \citep{lee2011seven, paradise2017creation}, to user feature oriented classifiers \citep{davis2016botornot, chu2012detecting, benevenuto2010detecting, wang2010detecting, egele2013compa, viswanath2014towards, thomas2011suspended}. All of these approaches treat accounts individually, but do not detect coordinated attacks. Extant work exists for coordinated attacks, a few instances of which are \emph{CopyCatch} for Facebook 'liked' pages \citep{beutel2013copycatch}, Twitter memes \citep{ratkiewicz2011detecting}, and more generally \emph{Sybil detection} in online communities \citep{benevenuto2009detecting, aggarwal2014data, cao2014uncovering, yang2014uncovering, ghosh2012understanding, tran2009sybil, yu2008sybillimit, danezis2009sybilinfer, yu2006sybilguard, wang2013you, alvisi2013sok, cao2012aiding}.   Our bot detection  algorithm is strongly inspired by the work of  \cite{zabih2004spatially} on image segmentation and the work  of \cite{marks2017building} on social network user geo-location.

\subsection{Opinions in Social Networks}
Key to assessing the impact of bots is understanding how they affect opinions in social networks.
A variety of models have been developed for the distribution of opinions in networks.
One of the earliest is the DeGroot model \citep{degroot1974reaching} where
users' opinions equal the weighted average of their neighbors' opinions.   This model has a similar flavor to many distributed consensus algorithms \citep{tsitsiklis1984problems,tsitsiklis1986distributed,olshevsky2009convergence,jadbabaie2003coordination}, as the goal of each user is to reach consensus
with his neighbors.  Related to the DeGroot model is the voter model  \citep{clifford1973model,holley1975ergodic} where users update their opinions
to match a randomly chosen neighbor.  There is a large body of theoretical research concerning the limiting behavior  in the voter model \citep{cox1986diffusive, gray1986duality, krapivsky1992kinetics, liggett2012interacting, sood2005voter}. 
Another class of models take a Bayesian perspective on how opinions evolve,
where each message a user posts causes his neighbors to update their belief
according to Bayes' theorem \citep{bikhchandani1992theory, banerjee2004word,acemoglu2011bayesian,banerjee1992simple,jackson2010social}.

The notion of stubborn users whose opinions do not evolve was introduced by \cite{mobilia2003does}.  Analysis has been done on the impact of stubborn users on opinions in networks \citep{galam2007role,wu2004social, chinellato2015dynamical, mobilia2007role,yildiz2013binary,acemouglu2013opinion,ghaderi2013opinion}.  
The model proposed by \cite{hunter2018opinion} is similar in flavor to the DeGroot model,
but is much more general, allowing users to grow stubborn with time at different rates and communicate noisy versions of their latent opinions.  Common to all of these models
is an opinion equilibrium where the non-stubborn users' opinions are determined
by the stubborn users.	\cite{vassio2014message} use this equilibrium to define harmonic influence centrality to measure the impact of individual nodes on the opinions in the network.

\section{Bot Behavior}\label{sec:botsbehavior}
Our first goal is to identify behavioral patterns of bots that distinguishes them from human users in social networks.
In this section we study the behavior of bots in the social network Twitter during several real world events.  We present our methods for data collection and bot labeling.  We then identify novel behaviors of the bots which we use to design our bot detection algorithm.

\subsection{Data Collection and Labeling}\label{sec:datasets_bots}
We collected Twitter data for six different  events that occurred in a variety of nations (U.S., France, Hungary), over various time periods (2015 to 2018), and for different durations.  Some of the events were elections in the U.S. and Hungary.  Others were for politically motivated conspiracy theories or scandals, such as Pizzagate and Macron Leaks.  Finally, there were activist groups such as Black Lives Matter (BLM) during 2015 and 2016.  We chose these events  because we suspected they would be targets for bots given their popularity and politically charged nature. Below we provide a brief background about these events.

\begin{enumerate}
	\item \textbf{Pizzagate}: During the 2016 US presidential election, WikiLeaks released the emails of John Podesta, who was Hillary Clinton's campaign manager. Conspiracy theorists claimed that some of those emails contained coded messages about human-trafficking and pedophilia rings run out of the basement of the Comet Ping Pong pizzeria located in Washington D.C.  The conspiracy spread through social media using the hashtag \#pizzagate.
	
	\item \textbf{Macron Leaks}:  The emails of candidate Emmanuel Macron were leaked during the 2017 French presidential election.  The emails leak spread rapidly on social media using the hashtag \#macronleaks.
	
	\item \textbf{2016 U.S. Presidential Debate}: This dataset consists of tweets by Twitter users who posted about the second debate of the 2016 U.S. presidential election  between Hillary Clinton and Donald Trump.
	
	\item \textbf{2018 Hungarian Parliamentary Elections}: The election took place in Hungary on the 8th of April 2018. It was viewed as a victory for the right-wing populist movement spreading through Europe.  
	
	\item \textbf{Black Lives Matter 2015}: The Black Lives Matter (BLM) movement was created in 2013 to protest police violence against the African-American community.  The movement went viral with the use of the hashtag $\#BlackLivesMatter$.  For this event we focus on social network data from 2015.
	
	\item \textbf{Black Lives Matter 2016}: This event focuses on social network data for Black Lives Matter from 2016. 
\end{enumerate}

We used Twitter's REST and Stream APIs \citep{ref:twitterSearchAPI} to collect user posts, known as \emph{tweets}, for the different events.  For the Pizzagate, BLM 2015 and Hungary Election datasets, we collected tweets containing relevant keywords.  For Pizzagate, the keyword was ``pizzagate'', for BLM 2015 the keywords were ``blm'' and ``blacklivesmatter'', and for the Hungary Election the keyword was ``HungarianElection''. The U.S. presidential debate dataset was provided in \cite{PDI7IN_2016}.  From this dataset we used all tweets posted by users who posted tweets about the second debate.
The Macron Leaks and BLM 2016 datasets were collected in \cite{macronLeaksDataSet,blm2016DataSet}.  The keywords used for the search criterion of these two datasets can be found within the provided references.  Details about our final datasets, including their size and dates covered, are provided in Table \ref{table:dataSets}.

\begin{table}[!hbt] \centering
	\caption{Time period, number of tweets, and number of unique users in the  Twitter datasets for different events. M is millions and K is thousands.}
	\label{table:dataSets}
	\centering
	\begin{tabular}{|l|l|c|}
		\hline
		Dataset     &  Time period  &  Number of tweets/users\\\hline
		Pizzagate &  Nov.-Dec. 2016 &  1.0M / 177K \\\hline
		BLM 2015  & Jan.-Dec. 2015 &  477K / 242K\\\hline
		U.S. presidential debate  & Jan.-Nov. 2016 & 2.4M / 78K \\\hline
		Macron leaks & May 2017 &   570K / 151K\\\hline
		Hungarian election & Apr. 2018 &  504K / 198K\\\hline
		BLM 2016 & Sep. 2016 &  1.3M / 546K \\\hline
	\end{tabular}
\end{table}

To obtain a ground truth for bot identities, we manually labeled approximately 300 Twitter accounts per dataset.  The accounts were randomly selected and were only required to have a minimum activity level.  Specifically, we focused on retweets, which is when a user shares a tweet posted by another user. We created a pool of the top 300 users retweeted by the highest number of distinct users  and a pool of the top 300 users who retweeted the highest number of distinct users. We then selected 150 accounts at random out of each of these two pools of users and merged them to create a test set to label. The reason we used this approach  is that bots represent a small fraction of the overall population (see Table \ref{table:data_hand}), hence selecting users at random out of the several hundred thousand uses per event came with a risk of having no bots in our test set. Focusing on the more active users increased the chances of collecting bots because we suspected that the bots would have elevated activity levels.

For each account, a human labeler was given three options: \emph{human}, \emph{bot}, or \emph{no idea}. We asked the labelers to focus on certain features of the accounts when selecting a label.  First was the account activity patterns, which included the number of retweets relative to the number of original content tweets, the raw volume of tweets, etc.  Bots are automated accounts, so they generally retweet others more than they tweet original content.  Second was the content of the tweets, the level of creativity, the presence of replies, and the diversity of topics discussed.  Bots may exhibit very simple language patterns because of their automation.  Third were other profile features such as the name, the profile and cover picture, and the  followers to friends ratio.  Bots may not have human profile pictures, use odd sounding names, and have a friend to follower ratio close to one because they are only followed by people they follow first.   We asked the labelers to combine these guidelines with their own sense about the accounts to apply the labels.  The number of accounts labeled as bots for each event are found in Table \ref{table:data_hand}.  Approximately 10\% of the accounts were labeled as bots across the different datasets.

\begin{table}[!hbt]
	\centering
	\caption{Number of Twitter accounts given bot labels and number of  Twitter accounts given any label   for each dataset.} \label{table:data_hand}	
	\begin{tabular}{|l|c|c|}
		\hline
		Dataset 		     &  Number of bot labels & Number of accounts labeled  \\	\hline
		Pizzagate	   		&  23  			 &  304 \\	\hline
		BLM 2015		   	   &  21 			&  262 \\	\hline
		U.S. presidential debate &  30 			& 300  \\	\hline 
		Macron leaks  	 & 19			  & 256  \\	\hline 
		Hungarian elections	& 24			 & 300  \\\hline 
		BLM 2016   & 30				& 285 \\   \hline
	\end{tabular}
\end{table}

\subsection{Retweet Graphs} \label{subsubsec:heterophily}
An important behavior in Twitter is known as \emph{retweeting}, which is when a user reposts a tweet to share it with his followers.  Retweeting is an easy way to promote a tweet and expose it to a larger audience.  A retweet is a tweet, but it can be viewed as a directed edge between the creator of the original tweet and the user retweeting it. The collection of all retweet edges for an event along with the users involved in the retweets constitute a \emph{retweet graph}.  We next examine properties of the retweet graphs for the datasets.

Because we labeled the users in the retweet graph as bot and human, there are four types of edges: bots retweeting bots, bots rewteeting humans, humans retweeting bots, and humans retweeting humans.  Figure \ref{fig:hetero_pizzagate} shows the Pizzagate retweet networks for these four types of edges.  In the figure the bots are located on the outer ring and the humans are located in the central ring. We see qualitatively that bots prefer to retweet humans instead of bots, and that humans prefer to retweet humans instead of bots. 
The  phenomenon where members of a group do not interact with each other, but do interact with members of different groups is known as \emph{heterophily} \citep{rogers1970homophily}, which is what the bots appear to be exhibiting in Figure \ref{fig:hetero_pizzagate}.  The humans exhibit the opposite phenomenon, known as \emph{homophily}, preferring to interact with each other rather than the bots. 
\begin{figure}[!h]
	\centering
	\includegraphics[scale=0.75]{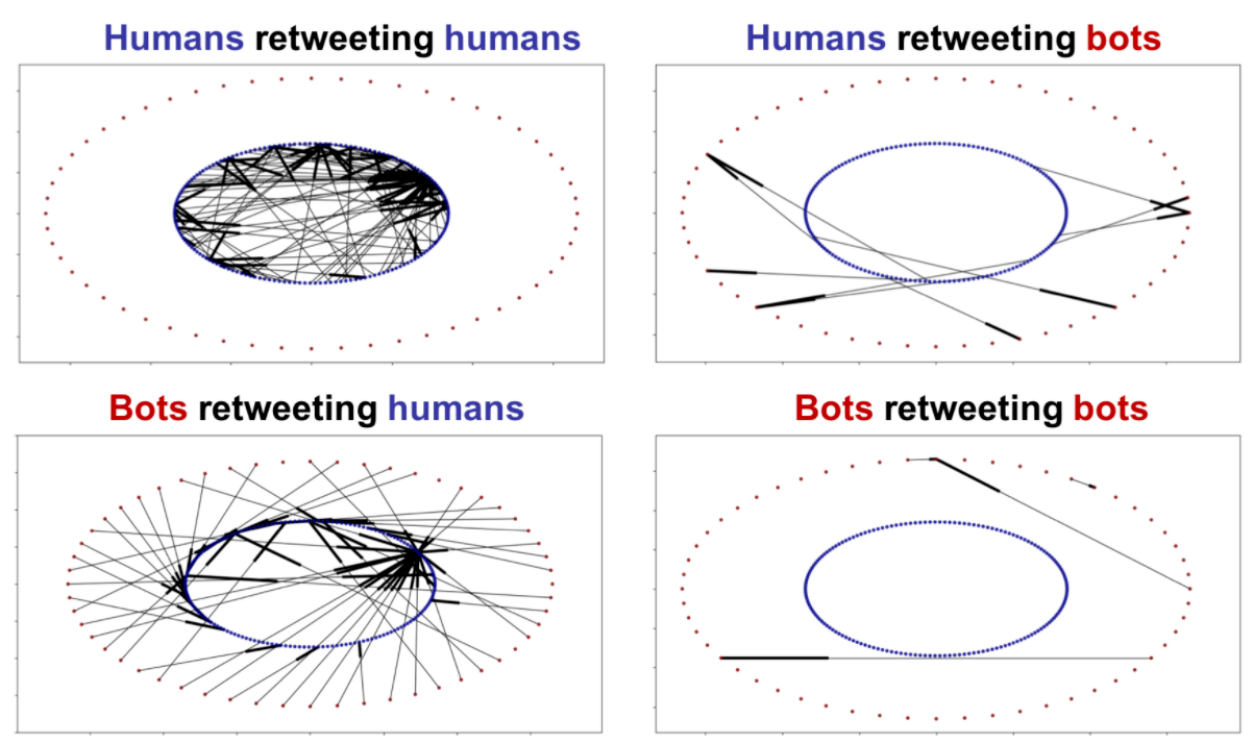} \\
	\caption{Retweet graphs for Pizzagate accounts labeled by humans.  Each network consists of edges between a certain pair of user types (bots and humans).} \label{fig:hetero_pizzagate}
\end{figure}

To obtain a more quantitative measure of the homophily and heterophily phenomena,
we broadened the set of ground truth labeled users by using  a popular machine learning based bot detection algorithm known as BotOrNot. \citep{davis2016botornot}.  This algorithm provides a probability of being a bot for a Twitter user.  We chose 0.5 as our threshold for being a bot. This process yielded labels for almost all users in each of the datasets, with the exception of a few hundred users that either got suspended or set their profiles to private.  Using BotOrNot gave us many more labels, but the reliability of these labels is not as high as the human labelers.  However, these noisy labels are sufficient for us to gain some understanding of the behavioral patterns in the data.

 The larger set of labeled users allowed us to look more closely at the different types of retweets.  For each account,
we calculated the total number of retweets it gave to humans divided by the number of unique humans it retweeted.  We did the same calculation for the bots it retweeted. This measures  the average number of retweets per human target and retweets per bot target for each account.  We refer to this quantity as \emph{retweets per target}.  If the bots exhibited heterophily, their retweets per target will be higher for the human targets.  Similarly, if the humans exhibited homophily, their retweets per target would also be higher for the human targets.

Figure \ref{fig:retweets_per_target} shows the retweets per target averaged over each user type for each dataset.  The bot heterophily and human homophily are evident from the plot.
To further quantify the difference, we performed a Kolmogorov-Smirnov (KS) test on the retweets per target distribution on each dataset.  We compared the distributions for bots retweeting bots to bots retweeting humans and humans retweeting humans to humans retweeting bots.  The p-value of each test is less than one percent, indicating that there is a statistically significant difference between the distributions.  This supports the hypothesis that there is heterophily for the bots and homophily for the humans.  We will use these properties to design our bot detection algorithm in Section \ref{sec:ising}.

\begin{figure}[!h]
	\centering
	\includegraphics[scale=0.6]
	{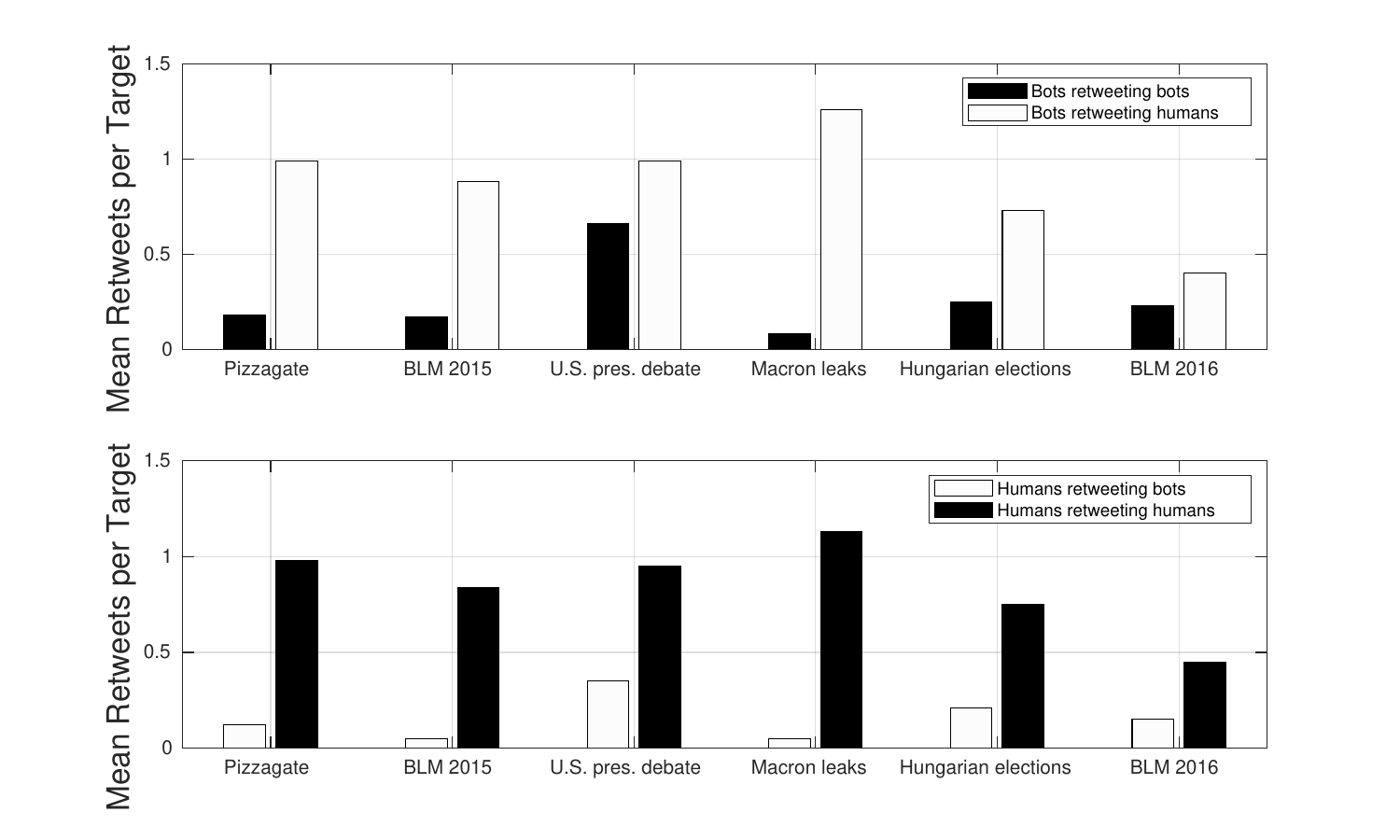} \\
	\caption{Plot of the mean retweets per target for humans and bots in the datasets.} \label{fig:retweets_per_target}
\end{figure}
%
%


\section{Ising Model Bot Detection Algorithm}\label{sec:ising}

In this section, we present our bot detection algorithm.  Contrary to machine learning algorithms such as BotOrNot \citep{davis2016botornot} which identify bots one at at time, our approach will be to simultaneously identify all bots in a retweet graph.  Our algorithm places a probability distribution on a graph and the labels of its nodes.  We represent this distribution using a graphical model, specifically the Ising model from statistical physics \citep{ising1925beitrag}.  Our approach to bot detection is inspired by work done in image segmentation \citep{zabih2004spatially} and network based geo-location \citep{marks2017building}.  

We begin with some notation.  Let $G = (V,E)$ denote a retweet graph with node set $V$ and edge set $E$.
For each  node $i\in V$ we observe features $\vx_{i}$ and for each pair of users $i,j\in V$ we observe interaction features $\vz_{ij}$.  The interaction features could be the number of retweets, out degree, and  in-degree of each node.   Each node $i$ in the graph has a latent variable $\Delta_i$ which is one if $i$ is a bot and zero otherwise. Our goal is to find the most likely configuration of the latent variables given the retweet graph and the features. 

We use the Ising model, which is a pairwise factor graph model, for the joint distribution of the latent variables given the observed features.  We define functions $\phi(\vx_{i},\Delta_{i})$ for each $i\in V$ and  $\psi(\vz_{ij},\Delta_{i},\Delta_{j})$ for each $i,j\in V$.
We refer to $\phi$ and $\psi$  as the \emph{node energy} and  \emph{link energy} functions.  For the Ising model, the joint distribution is determined by the energy of the latent variables.  Given a graph with node features $\mathbf X =\curly{\mathbf x_i}_{i\in V}$, interaction features $\mathbf Z =\curly{\mathbf z_{ij}}_{(i,j)\in E}$, and latent node labels $\mathbf\Delta = \curly{\Delta_i}_{i\in V}$, the Ising energy \citep{ising1925beitrag} is 
\begin{equation}
 E(\mathbf \Delta) = \sum_{i\in V}\phi(x_i,\Delta_i) + \sum_{(i,j)\in E}\psi(\mathbf z_{ij}, \Delta_i,\Delta_j)\label{eq:ising_energy}.
\end{equation}
The associated probability of the latent variables is
\begin{equation}
\Pr(\mathbf{\Delta}) =\frac{e^{-E(\mathbf \Delta)}}{\mathcal Z} \label{eq: joint_dist}
\end{equation}
where $\mathcal{Z}$ is the partition function.   
From this it can be seen that finding the maximum likelihood latent variable configuration reduces to minimizing the energy.  This is known to be NP-hard in general \citep{barahona1982computational}.  However, it has been shown that the inference problem is much easier if one specifies certain characteristics of the link and node energies \citep{zabih2004spatially, marks2017building}, which we do next.

\subsection{Link and Node Energies} \label{sec:energy}
\subsubsection{Link Energy}
We begin by defining the link energy functions.  First there is the case where there is no edge between nodes $i$ and $j$.  In this case we assume that we can infer very little about the latent variables $\Delta_i$ and $\Delta_j$, so  we set the link energy to be independent of the latent variables.  For simplicity, we assume that $\psi(\mathbf z_{ij},\Delta_i,\Delta_j)=0$ when there is no edge between $i$ and $j$.

For nodes $i$ and $j$ that have an edge $(i,j)$ between them,
we introduce the constants $\lambda_{10},\lambda_{00},\lambda_{11},  \lambda_{01}$ and a function $\psi_{ij}$, and set the link energies equal to
$$
\begin{array}{ll}
\psi(\vz_{ij},0,1)   &=  \lambda_{01}\psi_{ij} \\
\psi(\vz_{ij},1,1)   &=  \lambda_{11}\psi_{ij} \\
\psi(\vz_{ij},0,0)   &=  \lambda_{00}\psi_{ij} \\
\psi(\vz_{ij},1,0)   &=   \lambda_{10}\psi_{ij}.
\end{array}
$$\label{alambda}
We discuss in Section \ref{sec:constraints} how to determine the $\lambda$ constants in the above equations. 

To determine the link energy function $\psi_{ij}$ we use the following insight.    Suppose node $i$ retweets $z_i$ times (its out-degree),  and node $j$ receives  $z_j$ retweets (its in-degree), and $i$ retweets $j$ a total of $w_{ij}$ times.
We assume that if either degree is small, then the retweet edge from $i$ to $j$ provides little information about the node labels.  In simpler terms, the only edges that contain information are the ones where $j$ happens to be a popular target or  $i$ a suspiciously active retweeter.   Hence, whatever the link energies are, they should be insensitive to the labels of low degree nodes.  This suggests that $\psi_{ij}$ should approach zero as the degrees decrease.

Using this insight and following \cite{marks2017building} we define $\psi_{ij}$ as
\[
\psi_{ij} = \frac{\gamma w_{ij}}
{1+\exp(\alpha_{out}/z_i+\alpha_{in}/z_{j}-2)},
\]
where  $\gamma$ is a scaling factor that controls the weight of the link energy relative to the node energy, and  $\alpha_{out}, \alpha_{in}$ represent thresholds for the in-  and out-degrees, below which the link energy will be small. With this functional form, $\psi_{ij}$ increases monotonically as the in-degree and out-degree increase.  For degrees of zero, $\psi_{ij}$ is zero.  In this way, most information about the node labels is derived from higher degree nodes.

\subsubsection{Node Energy}  For the node energies, we must define $\phi(x_i,0)$ and $\phi(x_i,1)$ for a node with features $x_i$.  Because we can add a constant to the energy and not change the model, we set $\phi(x_i,0)=0$ for all $i\in V$.  We now have to select the node energy when the label is set to bot ($\Delta_i=1$).  We tried multiple options for this energy, but found that the most effective was to simply set it equal to zero.  That is, $\phi(x_i,0)=\phi(x_i,1)=0$ for all $i\in V$. What this is saying is that without any observed retweet edges, a node is equally likely to be a bot or a human.  While this approach does not incorporate any individual features of a user, we find that in practice it performs well with the added benefit of being quite simple to implement.  In Section \ref{sec:node_energy_robust} we show the performance of other choices for the node energy, including those that utilize information from other bot algorithms such as BotOrNot.  We find that the simple zero energy approach is better than or equal to these choices.


\subsection{Efficient Inference  via Minimum Cut}\label{sec:algo}
To find the most likely values for the labels  we want to minimize the Ising energy in equation \eqref{eq:ising_energy}.
It has been shown that minimizing the Ising energy can be done efficiently if the link energies have the properties given by the following result.
\begin{theorem}[\cite{kolmogorov2002energy}]
	Assume one is given a graph $G=(V,E)$ and associated Ising energy function given by equation \eqref{eq:ising_energy}.
	If the link energies satisfy
	\begin{equation}
	\psi(\vz_{ij},0,0)+\psi(\vz_{ij},1,1)\leq\psi(\vz_{ij},0,1)+\psi(\vz_{ij},1,0)\label{eq:submodularity}
	\end{equation}
	then the Ising energy function  is submodular  and can be minimized by solving a minimum cut problem.
\end{theorem}
To map the energy minimization to a minimum cut problem, we must define a new graph which we refer to as the \emph{energy graph}.  It is on this graph where solving a minimum cut problem provides the minimizer of the Ising energy.  

%

Figure \ref{fig:min-cut-graph} illustrates how to map a retweet graph into an energy graph.  The nodes of the energy graph are the nodes in the retweet graph plus a source node $s$ and a sink node $t$.  There are three types of edges in the energy graph.  For each node $i$ in the retweet graph, there is an edge from the source and an edge to the sink: $(s,i)$ and $(i,t)$.  There are also edges between every pair of nodes with an edge in the retweet graph.  

 For each node $i$ in the retweet graph,  every valid $s-t$ cut in the energy graph must either cut the edge $(s,i)$ or $(i,t)$.  If $(s,i)$ is cut, then $\Delta_i = 0$ and $i$ is a human.
Otherwise, $(i,t)$ is cut, $\Delta_i=1$ and $i$ is a bot.  This is how a cut in the energy graph maps to a label configuration.  By proper choice of the edge weights, the cut weight in the energy graph will equal the energy of the corresponding label configuration.  Then the minimum cut weight will provide the maximum likelihood configuration.

\begin{figure}[!h]
	\centering
	\includegraphics[scale=0.65]{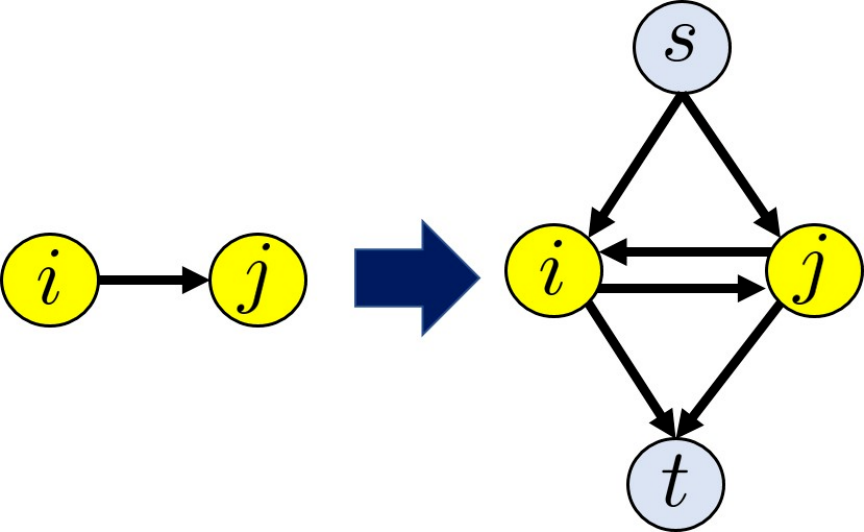} 
	\caption{An example retweet graph (left) and the corresponding energy graph (right).} \label{fig:min-cut-graph}
\end{figure}
We now define the edge
weights of the energy graph.  Denote the retweet graph by $G=(V,E)$.
For each edge $(i,j)\in E$ in the retweet graph, we add edges $(i,j)$ and $(j,i)$ to the energy graph with weights
\begin{align}
c_{(i,j)}   &= \frac{1}{2}\left(\psi(\vz_{ij},1,0) + \psi(\vz_{ij},0,1) - \psi(\vz_{ij},0,0) - \psi(\vz_{ij},1,1) \right)\nonumber\\
&=\frac{\psi_{ij}}{2}\paranth{\lambda_{10}+\lambda_{01}-\lambda_{00}-\lambda_{11}}\label{eq:cij}.
\end{align}
Note that these weights possess a symmetry, with $c_{(i,j)}=c_{(j,i)}$.
For each user $i\in V$, the weight of the edge $(s,i)$ in the energy graph is
\begin{align}
c_{(s,i)} =& \phi(\vx_{i},0) +\frac{1}{2}\sum_{j:(i,j)\in E}\psi(\vz_{ij},0,0)+
\frac{1}{4}\sum_{j:(i,j)\in E}(\psi(\vz_{ij},0,1)-\psi(\vz_{ij},1,0))+\nonumber\\
&	
+\frac{1}{2}\sum_{j:(j,i)\in E}\psi(\vz_{ji},0,0)+
\frac{1}{4}\sum_{j:(i,j)\in E}(\psi(\vz_{ji},1,0)-\psi(\vz_{ji},0,1))\nonumber\\
=&  \phi(\vx_{i},0)+\sum_{j:(i,j)\in E}\psi_{ij}\paranth{\frac{2\lambda_{00}+\lambda_{01}-\lambda_{10}}{4}}+
\sum_{j:(j,i)\in E}\psi_{ji}\paranth{\frac{2\lambda_{00}+\lambda_{10}-\lambda_{01}}{4}}.
\label{eq:csi}
\end{align}
For each user $i\in V$, the weight of the edge $(i,t)$ in the energy graph is
\begin{align}
c_{(i,t)} =& \phi(\vx_{i},1) +
\frac{1}{2}\sum_{j:(i,j)\in E}\psi(\vz_{ij},1,1)+
\frac{1}{4}\sum_{j:(i,j)\in E}(\psi(\vz_{ij},1,0)-\psi(\vz_{ij},0,1))+\nonumber\\
&	
+\frac{1}{2}\sum_{j:(j,i)\in E}\psi(\vz_{ji},1,1)+\frac{1}{4}\sum_{j:(i,j)\in E}(\psi(\vz_{ji},0,1)-\psi(\vz_{ji},1,0))\nonumber\\
=&  \phi(\vx_{i},1)+\sum_{j:(i,j)\in E}\psi_{ij}\paranth{\frac{2\lambda_{11}+\lambda_{10}-\lambda_{01}}{4}}+\sum_{j:(j,i)\in E}\psi_{ji}\paranth{\frac{2\lambda_{11}+\lambda_{01}-\lambda_{10}}{4}}.
\label{eq:cit}
\end{align}
With these edge weights, the weight of an $s-t$ cut
in the energy graph equals the energy of the corresponding label configuration in the interaction graph. 
We illustrate this using the example retweet graph in Figure \ref{fig:min-cut-graph}.  Consider the labels $\Delta_1=1$ and $\Delta_2=0$.  From equation \eqref{eq:ising_energy} the energy of this configuration is $\phi(x_1,1)+\phi(x_2,0)+\psi_{12}\lambda_{10}$.  The corresponding $s-t$ cut in the energy graph is $\curly{(1,t),(s,2),(1,2)}$.  Using the above expressions we find that the weights of the cut edges are
\begin{align*}
c_{(1,t)} & = \phi(x_1,1)+\frac{\psi_{12}}{4}\paranth{2\lambda_{11}+\lambda_{10}-\lambda_{01}}\\
c_{(s,2)} & = \phi(x_2,0)+\frac{\psi_{12}}{4}\paranth{2\lambda_{00}+\lambda_{10}-\lambda_{01}}\\
c_{(1,2)} & = \frac{\psi_{12}}{2}\paranth{\lambda_{10}+\lambda_{01}-\lambda_{00}-\lambda_{11}}.\\
\end{align*}
It can easily be checked that the weight of the cut equals the energy of the configuration.

\subsubsection{Link Energy Constraints} \label{sec:constraints}
 Recall that for a pair of nodes $i$ and $j$ connected by an edge $(i,j)$ and with labels $\Delta_{i}$ and $\Delta_{j}$, the corresponding link energy is $\psi(\mathbf z_{ij},\Delta_i,\Delta_j) = \lambda_{\Delta_i\Delta_j}\psi_{ij}$.
  For simplicity we set $\lambda_{01}=1$. The remaining three $\lambda$ parameters cannot be chosen arbitrarily. They must satisfy three types of constraints. First, they must respect the bot heterophily and human homophily properties  discussed in Section \ref{subsubsec:heterophily}.  This imposes the following constraints on the link energies when there is an edge from $i$ to $j$:
\begin{equation}
0 \leq \lambda_{10} \leq \lambda_{00} \leq \lambda_{11} \leq \lambda_{01}=1   \label{item:ineq_heterophily}.
\end{equation}
  These constraints simply say that a bot retweeting a human is more likely than a human retweeting a human, which is more likely that a bot retweeting a bot, which is more likely than a human retweeting a bot. 
We assume that humans make a conscious decision to retweet while bots are likely coded to retweet humans without much attention paid to the content.  Therefore, it is more likely for a bot to retweet a random human than for a human to retweet a random human. We also assume that humans will recognize bots and will most likely not retweet them.  Therefore, a bot retweeting a bot is more likely than a human retweeting a bot.  Though we did not do a rigorous analysis to support these assumptions, we will see that they produce a highly effective bot detection algorithm.

Second, in order to minimize the energy using a minimum cut, the parameters must satisfy the submodularity property from equation \eqref{eq:submodularity}.  This gives
\begin{equation}
\lambda_{10}+1 \geq \lambda_{00} + \lambda_{11}.  \label{item:ineq_submodularity}
\end{equation}
   Finally, because we are minimizing the energy using minimum cut, we need the edge weights in the energy graph to be non-negative. To satisfy this property, we set each term inside the summations in equations \eqref{eq:cij}, \eqref{eq:csi}, and \eqref{eq:cit}  to be greater than or equal to zero.  For the $c_{ij}$ weights this reduces to the submodularity constraint in equation \eqref{item:ineq_submodularity}.  For the $c_{si}$ weights this gives two inequalities: $2\lambda_{00}-\lambda_{10}+1\geq 0$, which is true by equation \eqref{item:ineq_heterophily}, and
$2\lambda_{00}+\lambda_{10}-1\geq 0$.  For the $c_{it}$ weights this gives $2\lambda_{11}-\lambda_{01}+1\geq 0$, which is also true by equation \eqref{item:ineq_heterophily}, and
$2\lambda_{11}+\lambda_{10}-1\geq 0$ which is satisfied if $2\lambda_{00}+\lambda_{10}-1\geq 0$ is satisfied.  Therefore, non-negative edge weights requires only one additional constraint:
\begin{equation}
2\lambda_{00} + \lambda_{10}-1\geq 0. \label{item:ineq_edgePos}
\end{equation}

The constraints in equations \eqref{item:ineq_heterophily}, \eqref{item:ineq_submodularity}, and \eqref{item:ineq_edgePos} constrain the $\lambda$ parameters to a polyhedron, but do not uniquely determine their values.     We will show a simple, yet effective approach to fully specify these values in Section \ref{sec:parameters}.

\section{Bot Detection Results}\label{sec:results_ising}
In this section, we present the results of the performance of the Ising model bot detection algorithm.  We first discuss how to select the values of the algorithm parameters.  Then we compare the performance of our algorithm to the state of the art BotOrNot bot detection algorithm   \citep{davis2016botornot}. 

\subsection{Algorithm Parameter Selection}\label{sec:parameters} 
To apply the Ising model algorihthm, we must choose values for  $\alpha_{in}$, $\alpha_{out}$, $\gamma$, and the $\lambda$ parameters.  We now present our approach to determine these values.  In addition, we performed tests  to show that the algorithm performance is robust to the precise parameter values.  These robustness checks are provided in Section \ref{sec:robust_ising}.

The $\alpha_{in}$ and $\alpha_{out}$ parameters relate to the degree of the nodes in a retweet graph.   We found that setting the values of $\alpha_{in}$ and $\alpha_{out}$ equal to an upper percentile of the in- and out-degree distributions worked well.  Using an upper percentile provides a  sense of the value at which we transition from a reasonable to an unusual number of retweets.  We select values that are close to the 99th percentile of the distributions.  The precise value is not important, as we show in Section \ref{sec:robust_ising}.

Since we set the node energies equal to zero, the Ising energy ends up being proportional to $\gamma$.  The node classification achieved with a minimum cut will not be affected by the value of $\gamma$. Therefore we simply set $\gamma = 1$.

The $\lambda$ parameters are constrained by the inequalities from Section \ref{sec:constraints}.  To further constrain the parameters, we make the submodularity inequality in equation \eqref{item:ineq_submodularity} an equality: $\lambda_{10} = \lambda_{00}+\lambda_{11}-1 $.  This sets $\lambda_{10}$ equal to its lower bound, which essentially says that the probability of the (1,0) configuration (bot retweeting human) is as different as possible from the other three configurations.  Substituting this expression for $\lambda_{10}$ into equation \eqref{item:ineq_edgePos} gives $\lambda_{11}\geq -3\lambda_{00}+2$.  This constraint combined with the heterophily/homophily constraint (equation \eqref{item:ineq_heterophily}) forms a polygon within which $\lambda_{00}$ and $\lambda_{11}$ are constrained, as illustrated in Figure \ref{fig:feasibleRegion}.  We choose the centroid of this polygon located at  $(\lambda_{00},\lambda_{11}) = (0.61,0.83)$ as the values of these two parameters.  Plugging this into the lower bound on $\lambda_{10}$ fully specifies all the $\lambda$ parameters as $(\lambda_{10},\lambda_{00},\lambda_{11},\lambda_{01}) = (0.44,0.61,0.83,1)$.

\begin{figure}[!h]
	\centering
	\includegraphics[scale=0.6]{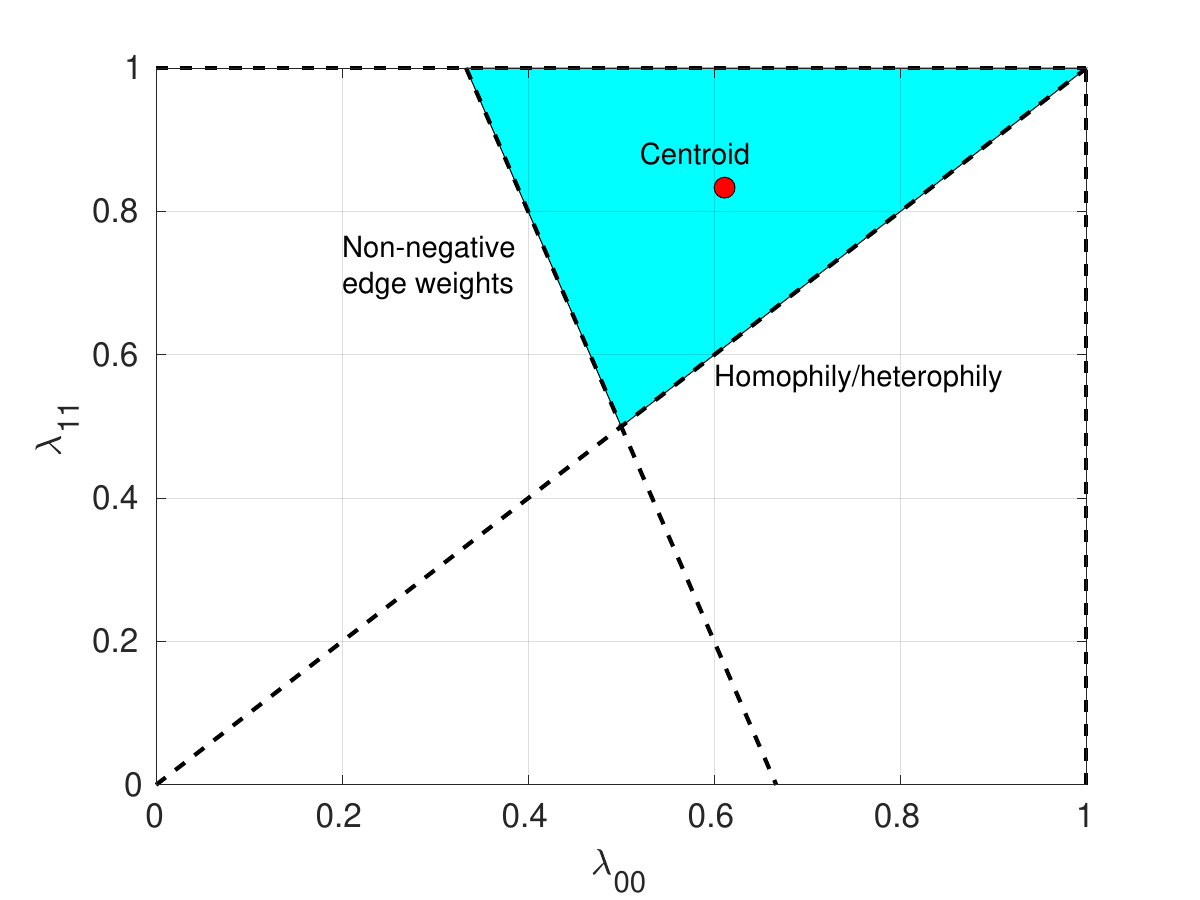} 
	\caption{Feasible region for $\lambda_{00}$ and $\lambda_{11}$ when $\lambda_{10}$ is set to its lower bound.  The sides of this polygon are labeled with the  constraint they enforce.  The centroid of the feasible region is $(\lambda_{00},\lambda_{11}) = (0.61,0.83)$. } \label{fig:feasibleRegion}
\end{figure}

\subsection{Algorithm Performance}\label{sec:roc_ising}
We now evaluate the performance of the Ising model algorithm for bot detection on the six events for which we have ground truth labels.  Recall that these labels were provided by humans and were limited in number (approximately 300 labels per event).  To increase our ground truth labeled set, we also included all Twitter users with  verified accounts. These users have undergone robust identity checks from Twitter in order to verify their identities.  We labeled all of these users as human. This increased the number of ground truth labels by approximately one to two thousand per dataset.

To apply the algorithm we had to set all the parameter values.  As mentioned earlier, we set $\gamma=1$ and $(\lambda_{10},\lambda_{00},\lambda_{11},\lambda_{01}) = (0.44,0.61,0.83,1)$.  We set $(\alpha_{out}, \alpha_{in})=(100,100)$ for all events except BLM 2016.    For that dataset we set $(\alpha_{out},\alpha_{in})=(100,1000)$.   These values represent the values above the 90th percentiles of the respective degree distributions in the retweet graphs.

We compare the Ising model algorithm to the BotOrNot algorithm \citep{davis2016botornot}, which is one of the top bot detection algorithms.  BotOrNot  collects a user's public profile and hundreds of its public tweets and mentions using the Twitter API.  It then extracts about 1,200 features related to the user's profile, friends, social network structure, temporal activity patterns, language, and sentiment.  These features are then fed to a machine learning algorithm which calculates the probability that the user is a bot.  

The Ising model algorithm uses only the structure of the retweet graph as input.  It knows nothing else about the users and the content they post. Compared to the Ising model algorithm, BotOrNot uses much more data and a more complex model.   Therefore,  one would expect BotOrNot to have superior performance.  

We used receiver operating characteristic (ROC) curves to compare the bot detection algorithms.  Because the Ising model algorithm provides a binary label for the accounts, we cannot directly calculate an ROC curve.  Instead, we use the inferred labels to calculate the probability of being a bot for each account conditional on the other labels using equation \eqref{eq: joint_dist}.  Formally, given a retweet graph $G=(V,E)$ and a node $i\in V$, let $\Delta_{-i} = \curly{\Delta_j}_{j\in V, j\neq i}  $ represent the labels of all nodes except $i$.  Then the conditional probability of $i$ being a bot is
\begin{equation}
P(\Delta_i=1|\Delta_{-i})  =\frac{1}{
	1+\exp\left(
	\phi(\vx_{i},1)
	-\phi(\vx_{i},0)
	+\sum_{j\neq i}\left[
	\psi(\vz_{ij},1,\Delta_{j})
	-\psi(\vz_{ij},0,\Delta_{j})
	\right]
	\right)
}\label{eq:conditional_prob}
\end{equation}

We use these conditional probabilities to calculate ROC curves for the Ising model and BotOrNot algorithms. The resulting ROC curves are shown in Figure \ref{fig:ROCs}. We observe that the Ising model algorithm achieves a true positive rate above 60\% at low false positive rates near 5\%. At similar false positive rates BotOrNot cannot achieve a true positive rate above 20\%.  Therefore, we see that the Ising model algorithm can achieve superior operating points than BotOrNot. 

The area under the curve (AUC) metric for an ROC curve is another performance measure.  An AUC of one is perfect detection, while an AUC of 0.5 is pure random guessing.  We show the AUC values in Table \ref{table:runtimeVSaccuracy}.  As can be seen, the Ising model algorithm achieves AUC's greater than BotOrNot on all events except for BLM 2015.  However, the AUC is lower on this event than the other events for both algorithms, suggesting that bot detection was in general difficult for this event.  We note that this  is the earliest event in our dataset.  Therefore, it is possible the behavior of bots changed with time, and so the Ising model algorithm would not necessarily outperform BotOrNot in this earlier period.

\begin{figure}[!h]
	\centering
	\includegraphics[scale=0.3]{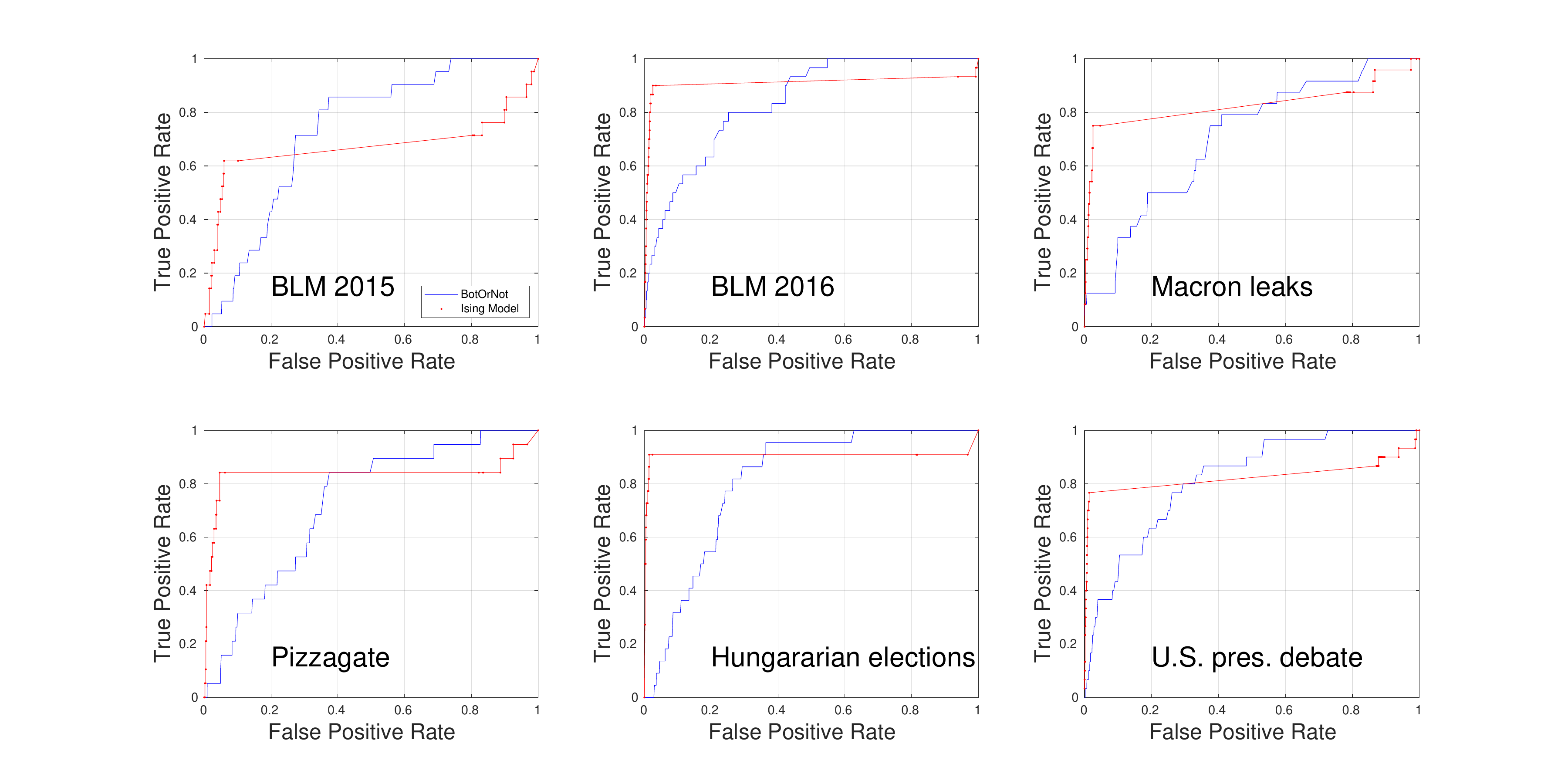} 
	\caption{Receiver operating characteristic (ROC) curves for the Ising model and BotOrNot bot detection algorithms on different Twitter datasets.} \label{fig:ROCs}
\end{figure}

\begin{table}[!hbt]
	\centering
	\caption{Area under the curve (AUC) values for the Ising model and BotOrNot bot detection algorithms on different Twitter datasets. Values in bold font represent the best of the two AUC scores on each dataset.} \label{table:runtimeVSaccuracy}
	\begin{tabular}{|l| c|c|}
		\hline
		Dataset &  AUC Ising & AUC BotOrNot\\ \hline
		Pizzagate &  \textbf{0.91}  &  0.81  \\ \hline 
		BLM 2015   &   0.67  &    \textbf{0.73}\\ \hline 
		U.S. presidential debate &  \textbf{0.83} &  0.82 \\  \hline
		Macron leaks    &  \textbf{0.84}  &  0.72\\ \hline 
		Hungarian elections    &  \textbf{0.83} & 0.71 \\ \hline 
		BLM 2016    &   \textbf{0.91}  &  0.84 \\
		\hline 
	\end{tabular}
\end{table}

\subsection{Evading Bot Detection}
Our bot detection algorithm relies upon the assumption of bot-human heterophily and human-human homophily.  This suggests potential ways for bot network operators to avoid detection.  The Ising model algorithm identifies users as bots if they are not being retweeted and  retweet others  often. 
For a bot to evade detection, it would need to be retweeted by other users.  Recall that we built the retweet graph using tweets about a topic of interest.  The bot can only be retweeted if it posts content of its own about this topic.  Therefore, to avoid detection the operator would need to have the bots post content about the topic.    If the bots do this and also retweet each other, then they will exhibit homophily and will appear more human to the Ising model algorithm.  However, this does require greater sophistication on the part of the bots.  Having the bots post content in addition to retweeting is more difficult to automate, but not impossible.  One simple way would be to simply copy the text of a tweet and have the bot post it as original tweet.   

If bots behaved in this more clever manner, the Ising model algorithm could still detect them if we modify how we define retweets.  Since bots would be copying tweets and posting them as their own rather than retweeting, we would need a method to detect this behavior in order to infer a retweet.
This could be done by measuring text similarity between and timing of tweets in order to infer duplication.  These inferred retweets would then form a graph to which the Ising model algorithm could be applied.  The manner in which we construct this inferred retweet graph would maintain bot-human heterophily because by identifying duplicated tweets, any retweets of bots would not count since they are not retweets of original tweets.  Therefore, we would expect the Ising model algorithm to be able to detect these more sophisticated bots.

\section{Assessing the Impact of Bots}\label{sec:assess}
The Ising model algorithm gives us a way to identify bots in social networks.  Our next goal is to assess the bots' impact on opinions in a social network. One way to do this is to ask what the opinions of users in the social network would have been if the bots were not present.  The difference in the mean opinion with and without the bots is a measure of their impact.  This counter-factual approach is commonly referred to as the potential outcomes framework \citep{rubin2005causal}.  One way to obtain this counter-factual condition would be to replicate the social network without bots.  Unfortunately we cannot do this.   However, if we had a model that allowed us to predict the opinions in the network with and without bots, we could use this to calculate opinions in the counter-factual condition and obtain the opinion shift due to the bots.

This model based approach is the one we will use to assess bot impact.  We face two challenges here.  First, we must determine what model should be used.  Second, we must find out how to use the model to calculate the opinions with and without the bots.  In Section \ref{sec:model} we present a model for opinion dynamics in a social network.  This model is very general in order to capture the heterogeneity present among social network users.  Then in Section \ref{sec:assess_method} we will construct a function based on this model to evaluate the bot impact.   The function will take as input a set of nodes and return the shift in the mean opinions due to the presence of the nodes.   Functions such as these which map nodes or sets of nodes to numerical values are referred to as \emph{network centralities}. The network centrality we develop is referred to as\textit{ generalized harmonic influence centrality}.  Sections \ref{sec:nn_measuring},  \ref{sec:data}, and \ref{sec:nn_training} show how to apply this centrality function to real social network datasets.  

\subsection{Opinion Dynamics Model}\label{sec:model}
We consider users in a directed social network or graph (we will use these terms interchangeably) $G=(V,E)$ with user set $V$ and edge set $E$.  Each user follows a set of individuals, which we refer to as his \emph{friends}.  If user $i$ follows user $j$, this is denoted by a directed edge $(j,i)$.  A user can see any social media  content posted by his friends.  To model the opinions in a social network we utilize the model proposed by \cite{hunter2018opinion} which is a generalization of the classic DeGroot model \citep{degroot1974reaching}.  We choose this model because it is quite rich and captures many of the behaviors exhibited by users in social networks.

Each user has an opinion on a binary issue.  For instance, a user can support or oppose a political candidate.  Let us define the opinion of a user $i$ at time $t$ as $\theta_i(t)$.  We assume the opinions are between zero and one.  If user $i$ posts at time $t$, the opinion of the post is $X_i(t)$, which is a random variable with expected value equal to $\theta_i(t)$ conditioned on $\theta_i(t)$.  This simply means the posts are unbiased representations of the user's current opinion.  

The model assumes that each user $i$ posts content according to a Poisson process of rate $\lambda_i$.   
The post of user $i$ shifts the opinions of his followers.  Formally, let $j$ be a follower of $i$.  When user $i$ posts content with opinion $X_i(t)$, user $j$ updates his opinion according to the following rule:
\begin{align}
\theta_j(t+1) & = (1-w_j(t))\theta_j(t) +w_j(t)X_i(t)\label{eq:update},
\end{align}
where $w_j(t)$ is a function that captures how stubborn user $j$ is.  As time increases, $w_j(t)$ approaches zero, indicating that users are becoming more stubborn, listening to their neighbors less and keeping their opinions constant.  Users are allowed to have $w_j(t)=0$ for all $t$.  We refer to these users as \emph{stubborn}, meaning that their opinions do not change.  These stubborn users could be hardened partisans who cannot be persuaded. They could also be bots which are programmed to post certain types of content.  

It was shown by \cite{hunter2018opinion} that in this model the user opinions reach an equilibrium for fairly general stubbornness functions $w_j(t)$.  In this equilibrium,  the opinion of a non-stubborn user $i$ is given by
\begin{align}
\sum_{j\in\text{friends of~}i} \lambda_i(\theta_i-\theta_j)=0\label{eq:equilibrium}.
\end{align}
Note that in the above expression the sum runs over both stubborn and non-stubborn users.  Another way to write the the equilibrium is to define $\mathbf \Psi$ as the vector of stubborn opinions and $\mathbf \theta$ as the vector of non-stubborn opinions.  Also define $V_0\subseteq V$ as the set of stubborn users and $V_1=V\setminus V_0$ as the set of non-stubborn users.  Then the equilibrium condition can be written in matrix form as
\begin{equation}
\mathbf G \mathbf\theta = \mathbf F\mathbf \Psi,\label{eq:equilibrium_matrix}
\end{equation}
where the matrix $\mathbf G$ is given by
\begin{align*}
\mathbf G_{ij} = \begin{cases}
-\sum_{k\in\text{friends of~}i} \lambda_{k}& \quad i=j, i\in V_1 \\
\lambda_{j} & \quad i\neq j, (j,i)\in E, i,j \in V_1 \\
0 & \text{else}, \\
\end{cases}
\end{align*}
and the matrix $\mathbf F$ is given by
\begin{align*}
\mathbf F_{ij} = \begin{cases}
\lambda_{j} & \quad (j,i)\in E, i\in V_1, j\in V_0  \\
0 & \text{else}. \\
\end{cases}
\end{align*}
The matrix form of the equilibrium condition highlights the fact that the non-stubborn opinions are linear combinations of the stubborn opinions.  It also shows that a unique equilibrium only exists if the matrix $\mathbf G$ is invertible.  In simple terms, the invertibility condition means that every non-stubborn user can be reached by at least one stubborn user.
  
There are less general instances of the model of \cite{hunter2018opinion} which have either deterministic communication, noiseless content opinions, or constant update weight functions \citep{degroot1974reaching, chatterjee1977towards, yildiz2013binary, acemouglu2013opinion, ghaderi2013opinion, vassio2014message}. However, all of these models reach the equilibrium given by equation \eqref{eq:equilibrium}, suggesting that this equilibrium may be a good model for how opinions are distributed in real social networks.

\subsection{Generalized Harmonic Influence Centrality for Assessing Bot Impact}\label{sec:assess_method}
The equilibrium condition in equation \eqref{eq:equilibrium} can be used to assess the impact of a set of users on the opinions in a network.  This was done  for individual users by \cite{vassio2014message} who defined the notion of \textit{harmonic influence centrality} as follows.  Assume one is given a network with stubborn nodes, non-stubborn nodes, and a node of interest $i$ which is also stubborn.  First, set the opinion of all stubborn nodes to zero except $i$, whose opinion is set equal to one.  In terms of equation \eqref{eq:equilibrium_matrix}, set $\mathbf\Psi_j=0$ for all $j\in V_0\setminus i$ and $\mathbf\Psi_i=1$.  Then use equation \eqref{eq:equilibrium} or \eqref{eq:equilibrium_matrix} to calculate  the  non-stubborn equilibrium opinions in the network. The harmonic influence centrality of $i$  is defined as mean of these non-stubborn  opinions.  This is also equal to the shift in the mean non-stubborn opinions caused by $i$ changing its opinion from zero to one because all other stubborn nodes have an opinion equal to zero.


Harmonic influence centrality provides one way to assess the impact of a single node in a network in terms of its ability to shift opinions.  However, it has a few drawbacks that make it not appropriate for real social networks.  First, it only measures impact for a single node.  In practice we would like to assess the impact of multiple nodes, for instance a group of bots.  Second, the opinions of stubborn nodes is set to zero, whereas in reality these opinions can be any value.  A more useful measure of impact would use the actual opinions of stubborn nodes. 

We modify harmonic influence centrality to address these drawbacks and produce a more effective way to assess the impact of multiple users or nodes in a real social network.
 To do this, we define generalized harmonic influence centrality as follows.
\begin{definition}
Let $G=(V,E)$ be a graph with stubborn node set $V_0\subseteq V$ and non-stubborn node set $V_1\subseteq V$.  For a set of nodes $S\subseteq V$, let $\theta$  and $\theta'$ be the vector of equilibrium opinions of the non-stubborn nodes given by equation \eqref{eq:equilibrium_matrix} with the nodes in $S$ included in $G$ and removed from $G$, respectively.  Then the \emph{generalized harmonic influence centrality} of $S$ is 
\begin{align}
\Delta(S) = \frac{1}{|V_1\setminus S|}\sum_{i\in V_1\setminus S} (\theta_i-\theta'_i).\label{eq:shift}
\end{align}
\end{definition}
Like harmonic influence centrality, generalized harmonic influence centrality takes into account the activity levels of the nodes in $S$ and the overall network structure.  What makes generalized harmonic influence centrality more useful for real social networks is its use of the actual opinions of stubborn users.  The generalized harmonic influence centrality of a set of nodes provides a more accurate assessment of their impact on the actual opinions in a network. The one difficulty with this measure is knowing the stubborn users' identities and opinions.  It is not clear how to determine the numerical value for a user's opinion in a social network, and even less clear how to determine who is stubborn.   We now show how to accomplish both of these tasks and make generalized harmonic influence centrality a useful operational tool.
 

\subsection{Neural Network for Measuring Opinions} \label{sec:nn_measuring}
Based upon the opinion dynamics model presented in Section \ref{sec:model}, we set the opinion of a user equal to the mean of the opinions of their tweets.  This follows from the assumption that the network has reached equilibrium and the tweets are unbiased with respect to the latent opinion.  The challenge is how to  estimate the opinion of the tweets. We do this using the neural network shown in Figure \ref{fig:CNN} which was proposed by  \cite{kim2014convolutional}.  The neural network takes as input the text of the tweet and outputs a score between zero and one which represents the opinion of the tweet with respect to the given topic.  Details on data processing and the neural network architecture are found in Section \ref{sec:neural_network}.  

The challenge of using the neural network is finding a sufficiently large set of training data which contains tweets labeled with ground truth opinions. One approach to obtain such a set  is to have human users manually look at a set of tweets and assign them labels based on their content.  This is a very time consuming task and would not allow us to obtain the large number of labeled tweets needed to train the neural network.  

To greatly enhance the size of our training data, we used the following approach. We assumed that a user's profile description contains very revealing information about his opinion.  For a given topic, we identified a set of hashtags and phrases that  indicate a strong opinion for or against the topic.  If a user's profile contained any of the phrases for or against the topic, we assumed that his latent opinion was one or zero. Furthermore, we labeled every one of his tweets about the topic with the same opinion  as his latent opinion.  This allowed us to quickly create huge datasets of labeled tweets that served as training data.  We show in Section \ref{sec:data} how we applied this approach to real social network data.  We find that the resulting neural networks obtained using this approach are quite effective at measuring tweet opinions.

\begin{figure}[h!]
	\centering
	\includegraphics[scale=1]{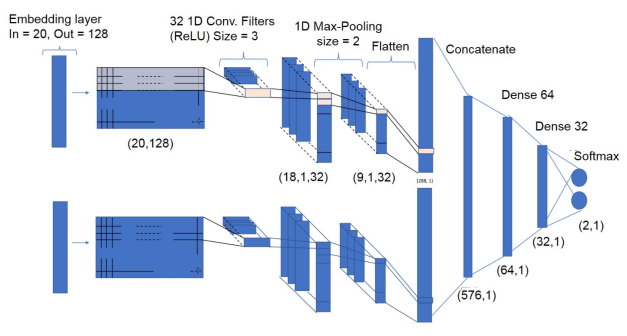}
	\caption{Diagram of the neural network architecture from \cite{kim2014convolutional} used to learn tweet opinions.  }
	\label{fig:CNN}
\end{figure}

\subsection{Dataset Description for Bot Impact Assesment}\label{sec:data}

We assess the impact of bots  on multiple Twitter datasets covering a variety of events.  These events were chosen because we suspected bots may be trying to influence the relevant social media discussion.  In this section we describe these datasets and details on training the neural network which is used to identify stubborn users and estimate their opinions. 
The datasets include tweets and also the follower graph formed by the users posting these tweets.  We used the Twitter API combined with a customized web crawler to collect all the edges of the follower graph for the users. Statistics about the datasets are shown in Table \ref{table:assess_data_stat}.  We now discuss details on the background and collection methodology for the datasets.
\begin{table}[!hbt] \centering
	\caption{Basic information about the Twitter datasets used to assess bot impact. M is millions and K is thousands.}
	\label{table:assess_data_stat}
	\centering
	\begin{tabular}{|l|l|c|c|c|}
		\hline
		Event     &    Data collection   & Number of  & Number of & Number of\\
		&    period            &  tweets & follower edges & users\\
		\hline
		U.S. presidential debate  & 	Jan.-Nov. 2016 & 2.4M & 5.4M & 78K \\ \hline
		Brexit 	 & Sep. 2018-Jan. 2019 &  18.5M  & 24.7M &105K \\ \hline
		Gilets Jaunes  & 	Jan.-Apr. 2019 & 2.3M & 4.6M & 40K \\
		\hline
	\end{tabular}
\end{table}  

\subsubsection{2016 U.S. Presidential Debate}
This dataset was previously described in Section \ref{sec:datasets_bots}.  As mentioned there, the 2016 U.S. presidential election has long been suspected of being attacked by bots. This dataset has 2.4 million tweets posted by 77,563 users.  The resulting follower graph contained 5.4 million edges. 

\subsubsection{Brexit}
The decision of the United Kingdom (U.K.) to leave the European Union on June 23, 2016, commonly referred to as \emph{Brexit}, is another event suspected of being influenced by social media bots \citep{ref:brexit_russia}.  In the years after the Brexit referendum, the U.K. government has been trying (unsuccessfully) to implement Brexit.  We focused on social network data during this period.  Specifically, we used the Twitter API to collect all tweets containing the word Brexit from September 27th, 2018 to January 31st, 2019. This resulted in tweets from 2.1 million users. We then selected a subset of 104,755 users who posted at least three tweets mentioning Brexit during the first two weeks of data collection.  These users had 18.5 million tweets and their follower graph had 24.7 million edges. 

\subsubsection{Gilets Jaunes}
Gilets Jaunes, or Yellow Vests, is a French populist movement that started in November 2018. Although it was initially a response to the sudden rise in fuel prices, it quickly became a generalized protest against the government of president Emmanuel Macron.  The protests have been going on every Saturday since November 2018, with each week being called a new ``Acte'' by the protesters.  

We collected Gilets Jaunes related tweets between January 26th, 2019 to April 29th, 2019 that contained any of the keywords shown in Table \ref{table:hashtag_choices_YellowVests}.  We needed to expand the keyword set beyond pure Gilets Jaunes words in order to collect a sufficient number of tweets for our analysis.  The resulting dataset contained 2.3 million tweets, 40,456 users, and 4.6 million edges in the associated follower graph.

\subsection{Neural Network Training}\label{sec:nn_training}
To label the tweets for training the neural network we needed to identify phrases and hashtags associated with extreme opinions for each dataset.  We identified these phrases by manually studying the language used in the social media discussion for each event.  The complete list of these phrases is provided in Section \ref{sec:labeling_tweets}.
We then identified all users who had these phrases in their Twitter profile description and labeled their tweets using the procedure outlined in Section \ref{sec:nn_measuring}.  Statistics for the resulting labeled datasets are summarized in Table \ref{table:training_set}.

We trained a different neural network for each dataset to learn the mapping from tweet text to tweet opinion.  For each event we trained on 80\% of the labeled data and tested on the remaining 20\%. The networks were trained using the deep learning library \emph{Keras} \citep{chollet2015}.  We used a cross-entropy loss function and trained over five epochs on a single CPU, resulting in a training time of a few hours.  Further details of the neural network  training process are  provided in Section \ref{sec:neural_network}.

On the testing data the neural network achieves an accuracy above 83\% for all of the datasets, as shown in Table \ref{table:training_set}.  This is quite a high accuracy and the result is even more impressive given that we used the same architecture for different languages (in this case English and French).  We show some example output opinions of the neural network for tweets from the datasets in Table \ref{table:exTweets_US}.  For each event, we show a tweet for, against, and neutral towards the topic.  As can be seen, the opinion estimates of the neural network align with the text of these tweets.  
\begin{table}[!hbt] \centering
	\caption{Training dataset descriptions and neural network performance. The first and second columns indicate the number of tweets used in the training set for each class. The third and fourth columns indicate the number of users who generated the training tweets.  The fifth column shows the accuracy of the neural networks on the testing data.  For each event, pro refers to pro-Trump for the U.S. presidential debate, pro-Brexit for Brexit, and pro-Gilets Jaunes for Gilets Jaunes.}
	\label{table:training_set}
	\centering
	\begin{tabular}{|c|c|c|c|c|c|}
		\hline
		Dataset & Number of &	Number of & Number of & Number of &Neural network  \\
		&pro-tweets&anti-tweets&pro-users&anti-users& accuracy \\\hline
		U.S. presidential debate & 100,000	&100,000		&	23,360 & 25,620 &92\%	 \\\hline
		Brexit 				& 								400,000				& 	400,000		&1,935	&	6,863	&86\%\\\hline
		Gilets Jaunes  		& 	130,000	& 	130,000		&  383   & 2,354 &83\%\\\hline
	\end{tabular}
\end{table} 

\begin{table}[h!]
	\begin{center}
		\caption{Tweets from testing data and their opinion scores given by the neural network for the  datasets.  An opinion of one  is pro-Trump for the U.S. presidential debate, pro-Brexit for Brexit, and pro-Gilets Jaunes for Gilets Jaunes.}
		\label{table:exTweets_US}
		\begin{tabular}{|c|l|c|} 
			\hline
			Dataset&Tweet&Neural network  \\
			&&opinion \\\hline
			U.S. pres. debate&@realDonaldTrump Your time is up I'm afraid, time & \\
				&to move on, your presidential campaign is untenable  & 0.33 \\\hline
			U.S. pres. debate&Hillary is struggling to defend herself. The group &  \\
				&simply doesn't believe her.  $\#$debate & 0.66 \\\hline
			U.S. pres. debate& I used to have a ton of respect for the & \\ 
				&Bush Family but if they can still vote for Clinton& 0.97\\
				&after all that has come out, my respect has 2 END & \\         \specialrule{2.5pt}{1pt}{1pt}

			Brexit& \#stopbrexit \#PeoplesVoter\#brexit \#Eunurses \#nurseshortage & 0.03 \\\hline
			Brexit& Britain will receive an economic boost on the back &\\ 
				&of a Brexit deal with the European Union, Philip Hammond &\\
				&has again claimed & 0.63  \\\hline
			Brexit& @Nigel\_Farage Wait for the remoaners to    make stupid comments&\\
				&of Russian interference on Brexit &  0.76\\        \specialrule{2.5pt}{1pt}{1pt}

			Gilets Jaunes & Il n'y a aucune raison que leurs revendications passent& \\ 
 				&avant d'autres, quelques dizaines de milliers repr\'esentant & \\
				&une minorit\'e ne vont pas d\'ecider pour la majorit\'e. & 0.0\\\hline
			Gilets Jaunes &\#Giletsjaunes \#Nancy Les manifestants ont r\`eussi \`a & \\
				&entrer dans le périmètre interdit dans le centre ville. &  0.5 \\\hline
			Gilets Jaunes &Aucun essoufflement pour l'\#ActeXV des \#GiletsJaune! & 0.85 \\\hline 	
	\end{tabular}
	\end{center}
\end{table}

\section{Results of Bot Impact Assessment}\label{sec:assessment_results}
We now present our results for assessing the impact of bots on different events using generalized harmonic influence centrality from Section \ref{sec:assess_method} and the trained neural networks described in Section \ref{sec:nn_training}.  Robustness results for our findings are provided in Section \ref{sec:robustness_tests}.

\subsection{Identifying Stubborn Users}
For each dataset we used the average of each user's tweet opinions determined by the neural network to obtain an estimate for their opinion.    
We use these values to identify stubborn users.
Studies have shown that stubborn users are likely to have very strong and extreme opinions \citep{martins2013building}. \cite{moussaid2013social} found that the majority of people will not change their opinion when their own confidence exceeds that of their partner.  These findings suggest that people with more extreme opinions are likely to be stubborn.  

To operationalize this notion of stubbornness, we chose $[0.0,0.1]$ and $[0.9,1.0]$ as stubborn intervals.  Any user whose opinion fell within one of these intervals was declared stubborn.   The number of stubborn users in each dataset for this choice of stubborn intervals   is shown in Table \ref{table:stubnotstub_stats}.   Recall that for our datasets, an opinion of one means pro-Trump, pro-Brexit, and pro-Gilets Jaunes.  As can be seen, there is an asymmetry in the stubborn users, with many more users in the lower stubbornness interval for each dataset.
The choice of stubbornness intervals is supported by the extant literature, yet is still somewhat arbitrary.  However,
  we show in Section \ref{sec:robustness_tests} that our results are robust to the precise choice of these intervals.    

\begin{table}[h!]
	\begin{center}
		\caption{Number of stubborn and non-stubborn users found in each dataset}
		\label{table:stubnotstub_stats}
		\begin{tabular}{|l|c|c|c|} 
			\hline
			Dataset & U.S. presidential debate &  ~Brexit~ &  Gilets Jaunes\\ \hline
			Number of non-stubborn users 	& 69,861&81,043 & 38,483  \\ \hline
			Number of stubborn users &  7,702 & 23,705 & 1,973\\ \hline
			Number of stubborn users in $[0.9,1.0]$   &  1,555 & 5,893 & 134\\ \hline
			Number of stubborn users in $[0.0,0.1]$   &  6,147 & 14,950 & 1,839\\ \hline		
		\end{tabular}
	\end{center}
\end{table}

We calculated the equilibrium opinions using equation \eqref{eq:equilibrium} and the opinions of the identified stubborn users. 
The posting rates of the users in this equation  were set to the number of their tweets in the dataset.  We were able to do this because all tweets were collected during the same time window and the equilibrium opinions are not changed by scaling of the rates. 

It is interesting to compare the non-stubborn opinions calculated using tweets and the neural network versus those calculated using the equilibrium condition.  The tweet based opinions are our ground-truth, and the equilibrium model is a prediction.  Moreover, the equilibrium model only utilizes the network structure and the opinions of the stubborn users who represent a small fraction of all users, as seen in Table \ref{table:stubnotstub_stats}.  The resulting equilibrium opinion statistics are shown in Table \ref{table:equilibrium_correlation}.  We find that the means of the equilibrium model are close to the tweet based opinions and there is a high correlation between the two sets of opinions.  While there is some error in the model predictions, it does appear that the equilibrium is capturing a significant aspect of the opinion distribution in the data.

\begin{table}[h!]
	\begin{center}
		\caption{Summary statistics of tweet based and equilibrium based non-stubborn opinions.}
		\label{table:equilibrium_correlation}
		\begin{tabular}{|c|c|c|c|} 
			\hline
			Dataset & Mean opinion  &
			          Mean opinion&
			          Correlation coefficient (p-value)\\
		& (tweet based)&  (equilibrium based) &  of tweet and equilibrium  opinions \\ \hline
		 U.S. presidential debate & 0.40 & 0.42 & 0.43 ($<10^{-6}$)\\ \hline
		 Brexit & 0.34 & 0.25 & 0.78 ($<10^{-6}$)\\ \hline
		 Gilets Jaunes & 0.52 & 0.41 & 0.78 ($<10^{-6}$)\\ \hline
		\end{tabular}
	\end{center}
\end{table}

\subsection{Bot Induced Equilibrium Shift}\label{sec:eq_shift}
We identified bots in each dataset using the Ising model algorithm.  Bots were determined by the output of the resulting minimum cut applied to the retweet graphs.  We show summary statistics of the detected bots in Table \ref{table:bot_count}.  As can be seen, the relative proportion of bots varies by dataset, but overall bots are a small fraction of the users.  There is also asymmetry in the bot opinions.  For the U.S. presidential debate, there are more pro-Trump than anti-Trump bots.  For Brexit, the anti-Brexit bots are dominant.  In Gilets Jaunes the bots are predominantly pro-Gilets Jaunes.  Based on these numbers, one can get a sense of how much of the discussion comes from bots and which side they support.  However, these numbers alone do not give us a clear sense of the impact of the bots because they ignore the bots' activity levels, connectivity, and network effects.

\begin{table}[h!]
	\begin{center}
		\caption{Number of Ising model bots in the upper and lower stubborn intervals for each dataset and the percentage of users  who are bots.}
		\label{table:bot_count}
		\begin{tabular}{|c|c|c|c|c|} 
			\hline
			Dataset & Number of& Percentage of users& Number of bots with   & Number of bots with  \\ 		
			&Ising model bots &  who are bots& opinion in $[0.0,0.1]$  & opinion in $[0.9,1.0]$ \\ \hline
			U.S. presidential &396& 0.5\% & 136 & 260 \\ 
			debate &&&& \\ \hline
			Brexit &5,854& 5.6\% & 3,931 & 1,923 \\ \hline
			Gilets Jaunes &4,874& 12.0\% & 1,491 & 3,383\\ \hline
		\end{tabular}
	\end{center}
\end{table}
To assess the bots' impact, we use generalized harmonic influence centrality and the measured stubborn user opinions.  We include the bots in the set of stubborn users.   The  mean non-stubborn opinions in each dataset with and without the bots are shown in Figure \ref{fig:bot_shift}.  The difference of these means is equal to the generalized harmonic influence centrality  of the bots and provides a measure of their impact.  Larger values of the  generalized harmonic influence centrality mean a larger opinion shift caused by the bots, and therefore larger impact.  We see that the shift varies by dataset.  In the U.S. presidential debate dataset the bots cause a large shift in the anti-Trump direction.  This is surprising given that the pro-Trump bots outnumber the anti-Trump bots.  In Brexit, the bots have nearly no effect.  In Gilets Jaunes the bots cause a large shift in the pro-Gilets Jaunes direction.

\begin{figure} 
	\centering
	\includegraphics[scale = .6]{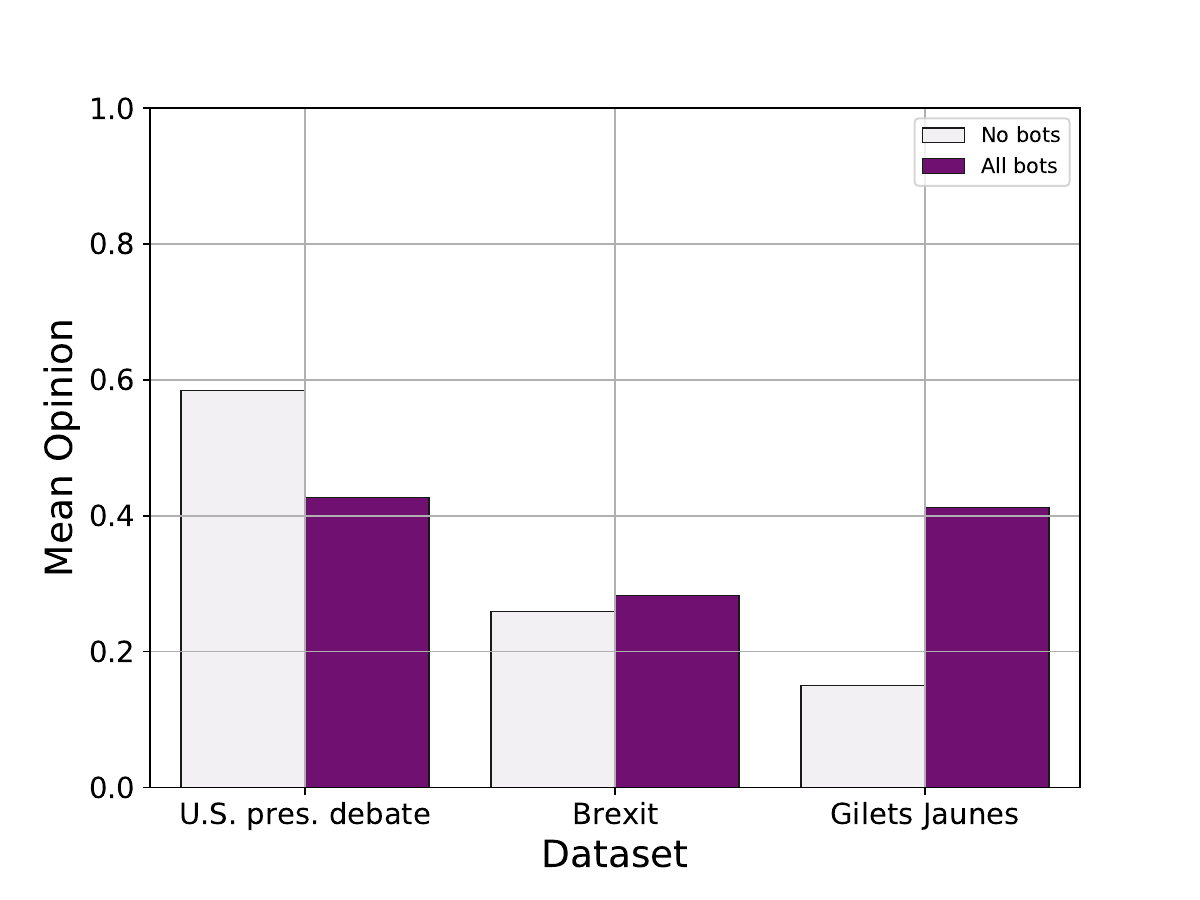}
	\caption{Bar graph of the mean non-stubborn equilibrium based opinion in each dataset with and without bots.  The shift in mean non-stubborn opinion caused by the bots is defined as their harmonic influence centrality.  The bots were identified using the Ising model algorithm.}
	\label{fig:bot_shift} 
\end{figure}

\subsection{Analysis of Bot Impact}
We see that the generalized harmonic influence centrality of the bots is different for each dataset.  It is interesting to look more closely at the data to try to understand what factors lead to the observed results.  From the equilibrium condition in equation \eqref{eq:equilibrium} we see that two factors which impact a user's opinion are his posting rate and the people he follows.  If a bot has a higher posting rate, it should have greater impact.  Similarly if a bot has more followers, it will also have a greater impact.  We now analyze these aspects in more detail to gain insights to the observed equilibrium opinion shifts.

\begin{table}[h!]
	\begin{center}
		\caption{Statistics of the posting rate distribution for the two bot classes in the datasets.  These classes are denoted by numeric intervals indicating where the bots' mean opinion lies.  Recall that for the datasets the $[0,0.5]$ interval corresponds to being anti-Trump, anti-Brexit, and anti-Gilets Jaunes.  Also shown is the p-value for a Kolmogorv-Smirnov (KS) test for differences in the distribution for the  bot classes.  The units for the posting rates are total tweets posted.}\label{table:rate}
		\begin{tabular}{|c|c|c|c|} 
			\hline
			Data set & $[0,0.5]$ bots posting rate  &$(0.5,1]$ bots posting rate& KS test \\
			&percentiles: 50\% (5\%, 95\%)  & percentiles:   50\% (5\%, 95\%)&p-value \\\hline
			U.S. presidential debate & 475 (49, 2245)& 318 (56, 1219)& $<10^{-3}$\\\hline
			Brexit & 864 (179, 4009)& 839 (181, 3447)& 1\\\hline
			Gilets Jaunes & 18 (9, 102)&  37 (11, 296)& $<10^{-6}$\\\hline
		\end{tabular}
	\end{center}
\end{table}

\begin{table}[h!]
	\begin{center}
		\caption{Statistics of the distribution of the follower count in the graph for the two bot classes in the datasets.  These classes are denoted by numeric intervals indicating where the bots' mean opinion lies.  Recall that for the datasets the $[0,0.5]$ interval corresponds to being anti-Trump, anti-Brexit, and anti-Gilets Jaunes.  Also shown is the p-value for a Kolmogorv-Smirnov (KS) test for differences in the distribution for the  bot classes.  }\label{table:follower}
		\begin{tabular}{|c|c|c|c|} 
			\hline
			Data set & $[0,0.5]$ bots follower count  &$(0.5,1]$ bots follower count  & KS test \\
			& percentiles: 50\% (5\%, 95\%)  &percentiles:  50\% (5\%, 95\%)&p-value \\\hline
			U.S. Presidential Debate & 54 (4, 275)& 46 (1, 264)& 0.23\\\hline
			Brexit & 299 (13, 2334)& 168 (13, 1025)& 1\\\hline
			Gilets Jaunes & 21 (1, 206)&  41 (4, 419)& $<10^{-6}$\\\hline
		\end{tabular}
	\end{center}
\end{table}

\subsubsection{U.S. Presidential Debate}
We found that the anti-Trump bots had a greater impact on the opinion equilibrium than the pro-Trump bots, despite being fewer in number.  This seems counter to what one would expect.  However, if we look at the posting rates of the bots in Table \ref{table:rate} we see evidence of why this is the case.  The distribution of the anti-Trump bots' posting rate has a larger median and also a heavier tail, as evidenced by the larger 95th percentile.  This heavier tail is visible in the cumulative distribution function (CDF) plots for the posting rates in Figure \ref{fig:bot_cdf_us}.  One can also see in this figure that both types of bots post much more frequently than the non-bot users.
  A Kolmogorov-Smirnov (KS) test shows that the bot posting rate distributions are statistically different.  Therefore, though being fewer in number, the anti-Trump bots are posting more frequently, which may be giving them the advantage.

We also looked at the distribution of the follower count in the graph for the two types of bots.  As seen in Table \ref{table:follower}, there is no statistical difference in these distributions.  Figure \ref{fig:bot_cdf_us}  shows the CDFs of the follower count in the graph for bots and non-bots.  The bots' follower count distributions clearly have a larger median and heavier tail than the non-bots, but there is no visible difference the distribution for the pro- and anti-Trump bots.  This suggests that the higher posting rate is the main factor in the equilibrium opinion shift.

To further support the rate hypothesis, we recalculated the equilibrium network opinions, but this time we gave every user the same posting rate.  The resulting mean opinions are shown in Table \ref{table:uniform_us}.  As can be seen, when the rates are equal,  the mean opinion shifts very slightly towards Trump.  In this case, the pro-Trump bots have an advantage, most likely because there are more of them.  Therefore it seems that the difference in the impact of pro and anti-Trump bots is due to the difference in their respective posting rate distributions.

\begin{figure} 
	\centering
	\includegraphics[scale = .35]{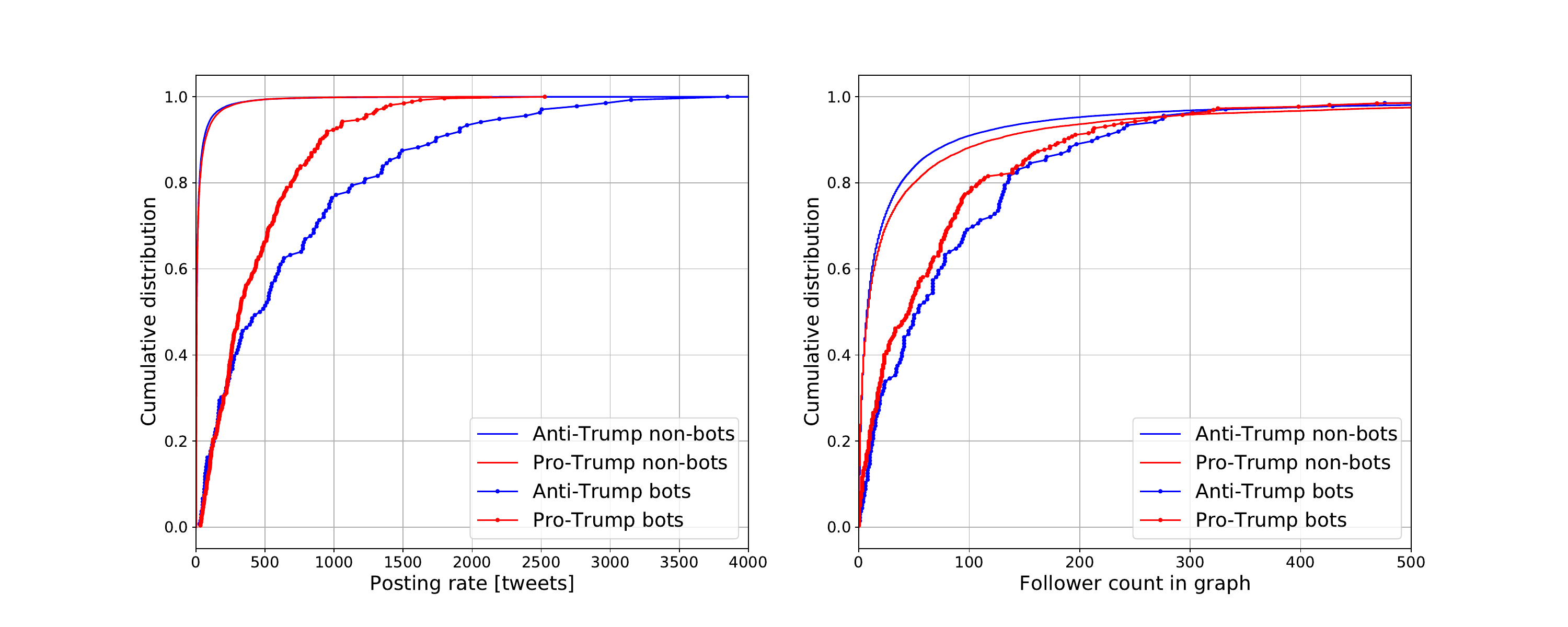}
	\caption{ Plots of the cumulative distributions of the bots' and non-bots' posting rate (left) and   follower count (right) in the U.S. presidential debate dataset. }
	\label{fig:bot_cdf_us} 
\end{figure}

\begin{table}[h!]
	\begin{center}
		\caption{Mean non-stubborn opinions in the U.S. presidential debate dataset with and without bots under actual posting rates and uniform posting rates.  The bots were identified using the Ising model algorithm. }\label{table:uniform_us}
		\begin{tabular}{|c|c|c|c|c|c|} 
			\hline
			 Posting rate&Mean opinion (no bots) & Mean opinion (all bots)\\\hline
			 Actual& 0.58 & 0.43\\\hline
			 Uniform& 0.23 &0.25\\\hline
		\end{tabular}
	\end{center}
\end{table}


\subsubsection{Brexit}
For Brexit there was a very small shift in the opinion mean.  We first look at the posting rate distributions.  We see in Table \ref{table:rate} that there is no statistical difference in the posting rate of the anti and pro-Brexit bots.  Figure \ref{fig:bot_cdf_brexit}  shows plots of the posting rate CDF's.  As can be seen, the bots post much more frequently than the non-bot users.  Therefore, one would expect the bots to have a non-trivial impact on the equilibrium opinions.  However, because bots on both sides have similar posting rates,  this is not the case.

We next look at the follower count of the bots.  From Figure \ref{fig:bot_cdf_brexit}  we see that the bots' follower count CDFs are higher than the non-bots, similar to  the bots in the U.S. presidential debate graph.  However, a KS test finds no statistical difference in the follower count distributions of both types of bots. 

We see that for Brexit that despite having higher posting rates than non-bot users, the bots do not cause any significant shift in the equilibrium opinion.  This is even more surprising given that the bots constitute 5.6\% of the network.  This is an order of magnitude larger than the percentage of bots in the U.S. presidential debate network.  This shows how simple measures of bot impact, such as their number, can be misleading when one ignores the structure of the network in which they exist.

\begin{figure} 
	\centering
	\includegraphics[scale = .35] {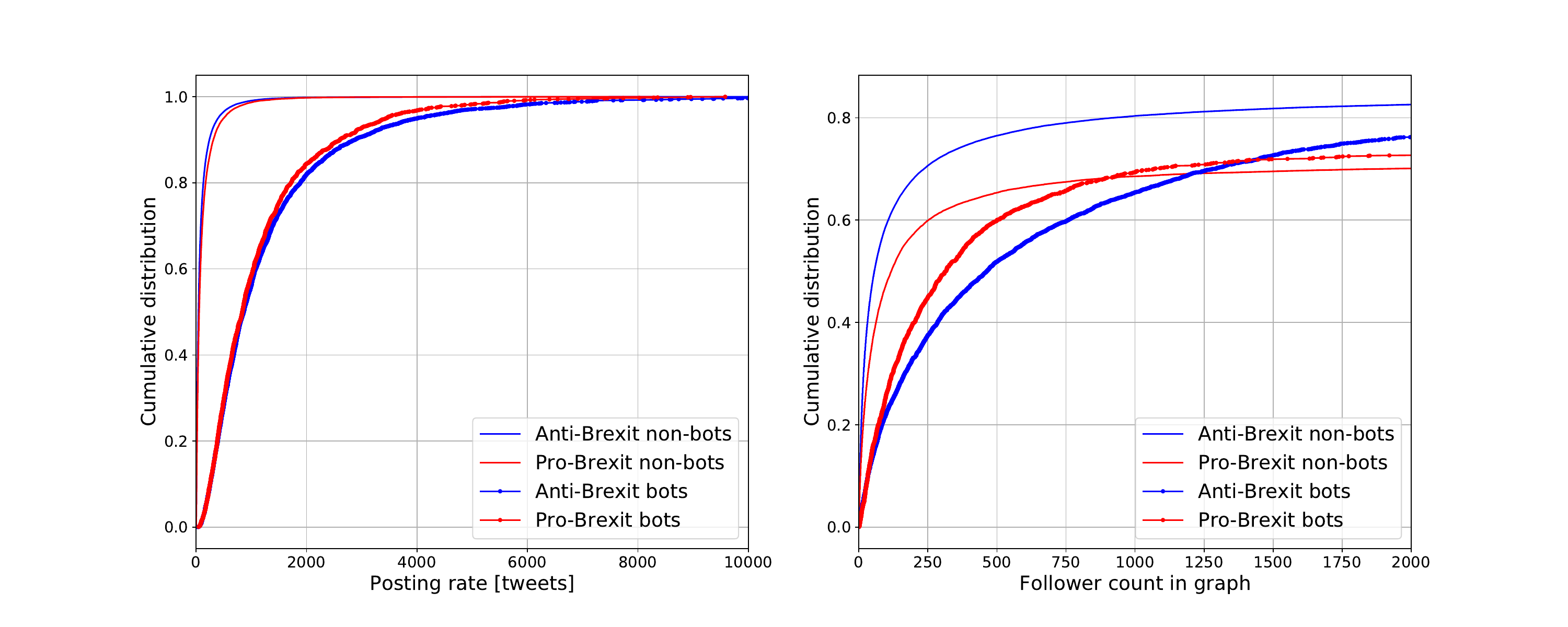}
	\caption{ Plots of the cumulative distributions of the bots' and non-bots' posting rate (left) and   follower count (right) in the Brexit dataset. }
	\label{fig:bot_cdf_brexit} 
\end{figure}


\subsubsection{Gilets Jaunes}
The pro-Gilets Jaunes bots had a strong impact on the equilibrium opinions.  Looking at the rate distributions in Table \ref{table:rate} we see that pro-Gilets Jaunes bots have a higher median posting rate and heavier tail than the anti-Gilets Jaunes bots. From the plots of the posting rate CDFs in Figure \ref{fig:bot_cdf_giletsjaunes} we see that anti-Gilets Jaunes users, both bot and non-bot, post at a higher rate than their opposition.  We do not see a clear distinction in the posting rate distribution of bots and non-bots.

Figure \ref{fig:bot_cdf_giletsjaunes} shows that the two classes of bots have very different in-graph follower count distributions.  This is further supported by a KS-test shown in Table \ref{table:follower}.  Therefore, it appears that for Gilets Jaunes, the shift comes from the elevated pro-Gilets Jaunes bot posting rates plus greater reach.  Over 12\% of the Gilets Jaunes graph is bots.  This is an instance where the bot count does suggest they have a large impact.  However, as we saw with Brexit, the count is not guaranteed to predict the shift in opinion equilibrium.

From detailed analysis of the datasets, we see that simple measures of impact can often be misleading. As we have seen in our analysis, there are instances with large numbers of active bots that have little impact.  Looking at dimensions such as posting rate or follower count alone is also not sufficient to assess bot impact.    The complex way these factors interact requires one to take into account the entire graph structure via a function such as generalized harmonic influence centrality.

\begin{figure} 
	\centering
	\includegraphics[scale = .35] {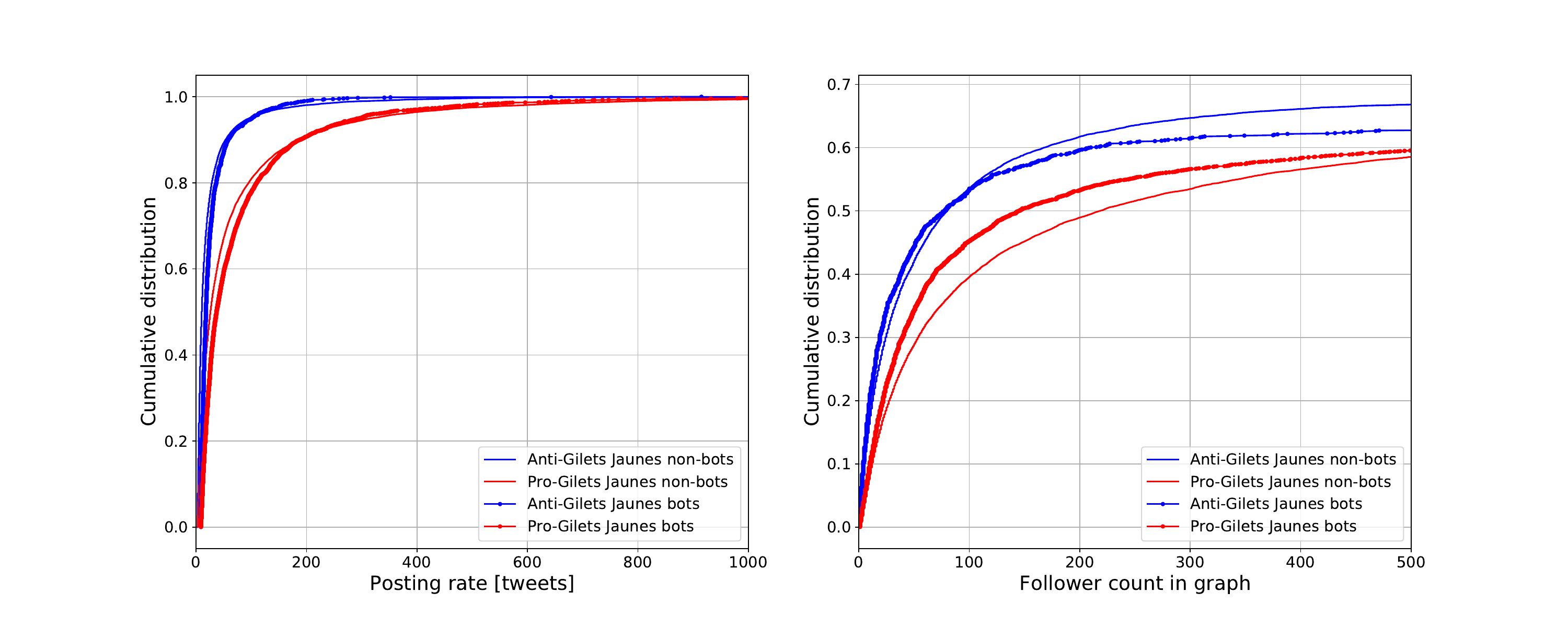}
	\caption{ Plots of the cumulative distributions of the bots' and non-bots' posting rate (left) and   follower count (right) in the Gilets Jaunes dataset. }
	\label{fig:bot_cdf_giletsjaunes} 
\end{figure}

\section{Future Work and Conclusion}\label{sec:conclusion}
\subsection{Future Work}
There are different possible directions for future work.  One interesting direction concerns a game-theoretic framework for bot detection and assessment.  Here we have looked at the problem from the perspective of the defender trying to identify bots and assess their impact.  However, the bot operator also faces interesting problems.  This operator wants to create bots that have impact and avoid detection.  However, there is a natural tradeoff.  If a bot is too active, then it can have a larger impact, but at the same time becomes easier to detect using our Ising model algorithm.  Therefore, there is likely an optimal activity level for the bots to balance impact with detection probability.  If the bot operator is using this type of strategy, then the defender may have a better way to detect the bots.  This suggests there may be equilibrium strategies in this game of bot operator versus defender.

There are also interesting theoretical questions with respect to the Ising model algorithm.  For instance, it would be useful to establish finite sample bounds for the performance of the algorithm.  These bounds would likely depend on not only the size of the rewteet graph or network, but also its structure.  More generally, one could investigate the relationship between algorithm performance and network structure in finite and infinite networks.  Perhaps there are networks where the algorithm performs well even under misspecification of the parameters.  If these networks resemble real social networks, this may even help explain the observed robustness of the algorithm.

\subsection{Conclusion}
Information operations require one to be able to assess the impact of influence campaigns in social networks.  This requires one to search for  bots conducting these campaigns, and then quantify the impact they have on the opinions in the social network.  Our work presents tools to accomplish both of these goals.  To identify the bots, we developed an algorithm based on the Ising model.  It uses minimal data and is able to jointly identify multiple bots with higher accuracy than state of art algorithms.  To assess the impact of opinions, we developed  generalized harmonic influence centrality which combined opinion dynamics models with neural networks.  This approach allows us to quantify the impact bots, or any set of users, have on the opinions in a social network.  Analysis on real datasets showed that the impact of bots varies, with some campaigns having minimal effect, while others cause large opinion shifts. 
The tools we developed here can be applied to multiple languages and social network types.  As the scale of propaganda campaigns increases from hostile actors, these information operations tools will find greater importance.

\ACKNOWLEDGMENT{This research was supported in part by the Office of Naval Research (ONR) and Charles Stark Draper Laboratory, Inc (Draper). The views presented here are those of the author and do not necessarily represent those of ONR, Draper, or MIT.}

\bibliography{Bots_Ising}

\begin{thebibliography}{100}
\providecommand{\natexlab}[1]{#1}
\providecommand{\url}[1]{\texttt{#1}}
\expandafter\ifx\csname urlstyle\endcsname\relax
  \providecommand{\doi}[1]{doi: #1}\else
  \providecommand{\doi}{doi: \begingroup \urlstyle{rm}\Url}\fi

\bibitem[Acemoglu et~al.(2011)Acemoglu, Dahleh, Lobel, and
  Ozdaglar]{acemoglu2011bayesian}
Daron Acemoglu, Munther~A Dahleh, Ilan Lobel, and Asuman Ozdaglar.
\newblock Bayesian learning in social networks.
\newblock \emph{The Review of Economic Studies}, 78\penalty0 (4):\penalty0
  1201--1236, 2011.

\bibitem[Acemo{\u{g}}lu et~al.(2013)Acemo{\u{g}}lu, Como, Fagnani, and
  Ozdaglar]{acemouglu2013opinion}
Daron Acemo{\u{g}}lu, Giacomo Como, Fabio Fagnani, and Asuman Ozdaglar.
\newblock Opinion fluctuations and disagreement in social networks.
\newblock \emph{Mathematics of Operations Research}, 38\penalty0 (1):\penalty0
  1--27, 2013.

\bibitem[Aggarwal(2014)]{aggarwal2014data}
Charu~C Aggarwal.
\newblock \emph{Data classification: algorithms and applications}.
\newblock CRC Press, 2014.

\bibitem[Alvisi et~al.(2013)Alvisi, Clement, Epasto, Lattanzi, and
  Panconesi]{alvisi2013sok}
Lorenzo Alvisi, Allen Clement, Alessandro Epasto, Silvio Lattanzi, and
  Alessandro Panconesi.
\newblock Sok: The evolution of sybil defense via social networks.
\newblock In \emph{Security and Privacy (SP), 2013 IEEE Symposium on}, pages
  382--396. IEEE, 2013.

\bibitem[Aral and Eckles(2019)]{aral2019protecting}
Sinan Aral and Dean Eckles.
\newblock Protecting elections from social media manipulation.
\newblock \emph{Science}, 365\penalty0 (6456):\penalty0 858--861, 2019.

\bibitem[Badawy et~al.(2018)Badawy, Ferrara, and Lerman]{badawy2018analyzing}
Adam Badawy, Emilio Ferrara, and Kristina Lerman.
\newblock Analyzing the digital traces of political manipulation: The 2016
  russian interference twitter campaign.
\newblock \emph{arXiv preprint arXiv:1802.04291}, 2018.

\bibitem[Banerjee and Fudenberg(2004)]{banerjee2004word}
Abhijit Banerjee and Drew Fudenberg.
\newblock Word-of-mouth learning.
\newblock \emph{Games and Economic Behavior}, 46\penalty0 (1):\penalty0 1--22,
  2004.

\bibitem[Banerjee(1992)]{banerjee1992simple}
Abhijit~V Banerjee.
\newblock A simple model of herd behavior.
\newblock \emph{The quarterly journal of economics}, 107\penalty0 (3):\penalty0
  797--817, 1992.

\bibitem[Barahona(1982)]{barahona1982computational}
Francisco Barahona.
\newblock On the computational complexity of ising spin glass models.
\newblock \emph{Journal of Physics A: Mathematical and General}, 15\penalty0
  (10):\penalty0 3241, 1982.

\bibitem[Bastos and Mercea(2019)]{bastos2019brexit}
Marco~T Bastos and Dan Mercea.
\newblock The brexit botnet and user-generated hyperpartisan news.
\newblock \emph{Social Science Computer Review}, 37\penalty0 (1):\penalty0
  38--54, 2019.

\bibitem[Benevenuto et~al.(2009)Benevenuto, Rodrigues, Almeida, Almeida, and
  Goncalves]{benevenuto2009detecting}
Fabricio Benevenuto, Tiago Rodrigues, Virgilio Almeida, Jussara Almeida, and
  Marcos Goncalves.
\newblock Detecting spammers and content promoters in online video social
  networks.
\newblock In \emph{Proceedings of the 32nd international ACM SIGIR conference
  on Research and development in information retrieval}, pages 620--627. ACM,
  2009.

\bibitem[Benevenuto et~al.(2010)Benevenuto, Magno, Rodrigues, and
  Almeida]{benevenuto2010detecting}
Fabricio Benevenuto, Gabriel Magno, Tiago Rodrigues, and Virgilio Almeida.
\newblock Detecting spammers on twitter.
\newblock In \emph{Collaboration, electronic messaging, anti-abuse and spam
  conference (CEAS)}, volume~6, page~12, 2010.

\bibitem[Bessi and Ferrara(2016)]{bessi2016social}
Alessandro Bessi and Emilio Ferrara.
\newblock Social bots distort the 2016 us presidential election online
  discussion.
\newblock \emph{First Monday}, 21\penalty0 (11-7), 2016.

\bibitem[Beutel et~al.(2013)Beutel, Xu, Guruswami, Palow, and
  Faloutsos]{beutel2013copycatch}
Alex Beutel, Wanhong Xu, Venkatesan Guruswami, Christopher Palow, and Christos
  Faloutsos.
\newblock Copycatch: stopping group attacks by spotting lockstep behavior in
  social networks.
\newblock In \emph{Proceedings of the 22nd international conference on World
  Wide Web}, pages 119--130. ACM, 2013.

\bibitem[Bikhchandani et~al.(1992)Bikhchandani, Hirshleifer, and
  Welch]{bikhchandani1992theory}
Sushil Bikhchandani, David Hirshleifer, and Ivo Welch.
\newblock A theory of fads, fashion, custom, and cultural change as
  informational cascades.
\newblock \emph{Journal of political Economy}, 100\penalty0 (5):\penalty0
  992--1026, 1992.

\bibitem[Boshmaf et~al.(2013)Boshmaf, Muslukhov, Beznosov, and
  Ripeanu]{boshmaf2013design}
Yazan Boshmaf, Ildar Muslukhov, Konstantin Beznosov, and Matei Ripeanu.
\newblock Design and analysis of a social botnet.
\newblock \emph{Computer Networks}, 57\penalty0 (2):\penalty0 556--578, 2013.

\bibitem[Byrnes(2016)]{byrnes2016bot}
Nanette Byrnes.
\newblock How the bot-y politic influenced this election.
\newblock \emph{Technology Rev.}, 2016.

\bibitem[Cao et~al.(2012)Cao, Sirivianos, Yang, and Pregueiro]{cao2012aiding}
Qiang Cao, Michael Sirivianos, Xiaowei Yang, and Tiago Pregueiro.
\newblock Aiding the detection of fake accounts in large scale social online
  services.
\newblock In \emph{Proceedings of the 9th USENIX conference on Networked
  Systems Design and Implementation}, pages 15--15. USENIX Association, 2012.

\bibitem[Cao et~al.(2014)Cao, Yang, Yu, and Palow]{cao2014uncovering}
Qiang Cao, Xiaowei Yang, Jieqi Yu, and Christopher Palow.
\newblock Uncovering large groups of active malicious accounts in online social
  networks.
\newblock In \emph{Proceedings of the 2014 ACM SIGSAC Conference on Computer
  and Communications Security}, pages 477--488. ACM, 2014.

\bibitem[Chatterjee and Seneta(1977)]{chatterjee1977towards}
Samprit Chatterjee and Eugene Seneta.
\newblock Towards consensus: Some convergence theorems on repeated averaging.
\newblock \emph{Journal of Applied Probability}, 14\penalty0 (1):\penalty0
  89--97, 1977.

\bibitem[Chinellato et~al.(2015)Chinellato, Epstein, Braha, Bar-Yam, and
  de~Aguiar]{chinellato2015dynamical}
David~D Chinellato, Irving~R Epstein, Dan Braha, Yaneer Bar-Yam, and Marcus~AM
  de~Aguiar.
\newblock Dynamical response of networks under external perturbations: exact
  results.
\newblock \emph{Journal of Statistical Physics}, 159\penalty0 (2):\penalty0
  221--230, 2015.

\bibitem[Chollet(2015)]{chollet2015}
François Chollet.
\newblock keras.
\newblock \url{https://github.com/fchollet/keras}, 2015.

\bibitem[Chu et~al.(2012)Chu, Gianvecchio, Wang, and Jajodia]{chu2012detecting}
Zi~Chu, Steven Gianvecchio, Haining Wang, and Sushil Jajodia.
\newblock Detecting automation of twitter accounts: Are you a human, bot, or
  cyborg?
\newblock \emph{IEEE Transactions on Dependable and Secure Computing},
  9\penalty0 (6):\penalty0 811--824, 2012.

\bibitem[Clifford and Sudbury(1973)]{clifford1973model}
Peter Clifford and Aidan Sudbury.
\newblock A model for spatial conflict.
\newblock \emph{Biometrika}, 60\penalty0 (3):\penalty0 581--588, 1973.

\bibitem[Cox and Griffeath(1986)]{cox1986diffusive}
J~Theodore Cox and David Griffeath.
\newblock Diffusive clustering in the two dimensional voter model.
\newblock \emph{The Annals of Probability}, pages 347--370, 1986.

\bibitem[Danezis and Mittal(2009)]{danezis2009sybilinfer}
George Danezis and Prateek Mittal.
\newblock Sybilinfer: Detecting sybil nodes using social networks.
\newblock In \emph{NDSS}, pages 1--15. San Diego, CA, 2009.

\bibitem[Davis et~al.(2016)Davis, Varol, Ferrara, Flammini, and
  Menczer]{davis2016botornot}
Clayton~Allen Davis, Onur Varol, Emilio Ferrara, Alessandro Flammini, and
  Filippo Menczer.
\newblock Botornot: A system to evaluate social bots.
\newblock In \emph{Proceedings of the 25th International Conference Companion
  on World Wide Web}, pages 273--274. International World Wide Web Conferences
  Steering Committee, 2016.

\bibitem[DeGroot(1974)]{degroot1974reaching}
Morris~H DeGroot.
\newblock Reaching a consensus.
\newblock \emph{Journal of the American Statistical Association}, 69\penalty0
  (345):\penalty0 118--121, 1974.

\bibitem[DoD(2012)]{dod2012joint}
US~DoD.
\newblock Joint publication 3-13.
\newblock \emph{Information Operations}, 2012.

\bibitem[Egele et~al.(2013)Egele, Stringhini, Kruegel, and
  Vigna]{egele2013compa}
Manuel Egele, Gianluca Stringhini, Christopher Kruegel, and Giovanni Vigna.
\newblock Compa: Detecting compromised accounts on social networks.
\newblock In \emph{NDSS}, 2013.

\bibitem[Elovici et~al.(2014)Elovici, Fire, Herzberg, and
  Shulman]{elovici2014ethical}
Yuval Elovici, Michael Fire, Amir Herzberg, and Haya Shulman.
\newblock Ethical considerations when employing fake identities in online
  social networks for research.
\newblock \emph{Science and engineering ethics}, 20\penalty0 (4):\penalty0
  1027--1043, 2014.

\bibitem[Fandos and Shane(2017)]{ref:russianbots_govtresponse}
Nocholas Fandos and Scott Shane.
\newblock {Senator Berates Twitter Over ‘Inadequate’ Inquiry Into Russian
  Meddling }.
\newblock \emph{The New York Times}, September 2017.
\newblock URL
  \url{https://www.nytimes.com/2017/09/28/us/politics/twitter-russia-interference-2016-election-investigation.html?mtrref=www.google.com}.

\bibitem[Ferrara(2017)]{ferrara2017disinformation}
Emilio Ferrara.
\newblock Disinformation and social bot operations in the run up to the 2017
  french presidential election.
\newblock 2017.

\bibitem[Ferrara et~al.(2016)Ferrara, Varol, Davis, Menczer, and
  Flammini]{ferrara2016rise}
Emilio Ferrara, Onur Varol, Clayton Davis, Filippo Menczer, and Alessandro
  Flammini.
\newblock The rise of social bots.
\newblock \emph{Communications of the ACM}, 59\penalty0 (7):\penalty0 96--104,
  2016.

\bibitem[Freitas et~al.(2015)Freitas, Benevenuto, Ghosh, and
  Veloso]{freitas2015reverse}
Carlos Freitas, Fabricio Benevenuto, Saptarshi Ghosh, and Adriano Veloso.
\newblock Reverse engineering socialbot infiltration strategies in twitter.
\newblock In \emph{Advances in Social Networks Analysis and Mining (ASONAM),
  2015 IEEE/ACM International Conference on}, pages 25--32. IEEE, 2015.

\bibitem[Galam and Jacobs(2007)]{galam2007role}
Serge Galam and Frans Jacobs.
\newblock The role of inflexible minorities in the breaking of democratic
  opinion dynamics.
\newblock \emph{Physica A: Statistical Mechanics and its Applications},
  381:\penalty0 366--376, 2007.

\bibitem[Ghaderi and Srikant(2013)]{ghaderi2013opinion}
Javad Ghaderi and R~Srikant.
\newblock Opinion dynamics in social networks: A local interaction game with
  stubborn agents.
\newblock In \emph{American Control Conference (ACC), 2013}, pages 1982--1987.
  IEEE, 2013.

\bibitem[Ghosh et~al.(2012)Ghosh, Viswanath, Kooti, Sharma, Korlam, Benevenuto,
  Ganguly, and Gummadi]{ghosh2012understanding}
Saptarshi Ghosh, Bimal Viswanath, Farshad Kooti, Naveen~Kumar Sharma, Gautam
  Korlam, Fabricio Benevenuto, Niloy Ganguly, and Krishna~Phani Gummadi.
\newblock Understanding and combating link farming in the twitter social
  network.
\newblock In \emph{Proceedings of the 21st international conference on World
  Wide Web}, pages 61--70. ACM, 2012.

\bibitem[Goldberg and Levy(2014)]{goldberg2014word2vec}
Yoav Goldberg and Omer Levy.
\newblock word2vec explained: deriving mikolov et al.'s negative-sampling
  word-embedding method.
\newblock \emph{arXiv preprint arXiv:1402.3722}, 2014.

\bibitem[Gray(1986)]{gray1986duality}
Lawrence Gray.
\newblock Duality for general attractive spin systems with applications in one
  dimension.
\newblock \emph{The Annals of Probability}, pages 371--396, 1986.

\bibitem[Guilbeault and Woolley(2016)]{guilbeault2016twitter}
Douglas Guilbeault and Samuel Woolley.
\newblock How twitter bots are shaping the election.
\newblock \emph{The Atlantic}, 1, 2016.

\bibitem[Han et~al.(2014)Han, Cook, and Baldwin]{han2014text}
Bo~Han, Paul Cook, and Timothy Baldwin.
\newblock Text-based twitter user geolocation prediction.
\newblock \emph{Journal of Artificial Intelligence Research}, 49:\penalty0
  451--500, 2014.

\bibitem[Holley and Liggett(1975)]{holley1975ergodic}
Richard~A Holley and Thomas~M Liggett.
\newblock Ergodic theorems for weakly interacting infinite systems and the
  voter model.
\newblock \emph{The annals of probability}, pages 643--663, 1975.

\bibitem[Hunter and Zaman(2018)]{hunter2018opinion}
D~Scott Hunter and Tauhid Zaman.
\newblock Optimizing opinions with stubborn agents under time-varying dynamics.
\newblock \emph{arXiv preprint arXiv:1806.11253}, 2018.

\bibitem[Hwang et~al.(2012)Hwang, Pearce, and Nanis]{hwang2012socialbots}
Tim Hwang, Ian Pearce, and Max Nanis.
\newblock Socialbots: Voices from the fronts.
\newblock \emph{interactions}, 19\penalty0 (2):\penalty0 38--45, 2012.

\bibitem[Ising(1925)]{ising1925beitrag}
Ernst Ising.
\newblock Beitrag zur theorie des ferromagnetismus.
\newblock \emph{Zeitschrift f{\"u}r Physik A Hadrons and Nuclei}, 31\penalty0
  (1):\penalty0 253--258, 1925.

\bibitem[Jackson(2010)]{jackson2010social}
Matthew~O Jackson.
\newblock \emph{Social and economic networks}.
\newblock Princeton university press, 2010.

\bibitem[Jadbabaie et~al.(2003)Jadbabaie, Lin, and
  Morse]{jadbabaie2003coordination}
Ali Jadbabaie, Jie Lin, and A~Stephen Morse.
\newblock Coordination of groups of mobile autonomous agents using nearest
  neighbor rules.
\newblock \emph{IEEE Transactions on automatic control}, 48\penalty0
  (6):\penalty0 988--1001, 2003.

\bibitem[Jurgens et~al.(2015)Jurgens, Finethy, McCorriston, Xu, and
  Ruths]{jurgens2015geolocation}
David Jurgens, Tyler Finethy, James McCorriston, Yi~Tian Xu, and Derek Ruths.
\newblock Geolocation prediction in twitter using social networks: A critical
  analysis and review of current practice.
\newblock \emph{Icwsm}, 15:\penalty0 188--197, 2015.

\bibitem[Kempe et~al.(2003)Kempe, Kleinberg, and Tardos]{kempe2003maximizing}
David Kempe, Jon Kleinberg, and {\'E}va Tardos.
\newblock Maximizing the spread of influence through a social network.
\newblock In \emph{Proceedings of the ninth ACM SIGKDD international conference
  on Knowledge discovery and data mining}, pages 137--146. ACM, 2003.

\bibitem[Kempe et~al.(2005)Kempe, Kleinberg, and Tardos]{kempe2005influential}
David Kempe, Jon Kleinberg, and {\'E}va Tardos.
\newblock Influential nodes in a diffusion model for social networks.
\newblock In \emph{Automata, languages and programming}, pages 1127--1138.
  Springer, 2005.

\bibitem[Kim(2014)]{kim2014convolutional}
Yoon Kim.
\newblock Convolutional neural networks for sentence classification.
\newblock \emph{arXiv preprint arXiv:1408.5882}, 2014.

\bibitem[Klausen et~al.(2018)Klausen, Marks, and Zaman]{klausen2018finding}
Jytte Klausen, Christopher~E Marks, and Tauhid Zaman.
\newblock Finding extremists in online social networks.
\newblock \emph{Operations Research}, 66\penalty0 (4):\penalty0 957--976, 2018.

\bibitem[Kolmogorov and Zabih(2002)]{kolmogorov2002energy}
Vladimir Kolmogorov and Ramin Zabih.
\newblock What energy functions can be minimized via graph cuts?
\newblock In \emph{European conference on computer vision}, pages 65--81.
  Springer, 2002.

\bibitem[Krapivsky(1992)]{krapivsky1992kinetics}
PL~Krapivsky.
\newblock Kinetics of monomer-monomer surface catalytic reactions.
\newblock \emph{Physical Review A}, 45\penalty0 (2):\penalty0 1067, 1992.

\bibitem[Lee et~al.(2011)Lee, Eoff, and Caverlee]{lee2011seven}
Kyumin Lee, Brian~David Eoff, and James Caverlee.
\newblock Seven months with the devils: A long-term study of content polluters
  on twitter.
\newblock In \emph{ICWSM}, 2011.

\bibitem[Liggett(2012)]{liggett2012interacting}
Thomas~Milton Liggett.
\newblock \emph{Interacting particle systems}, volume 276.
\newblock Springer Science \& Business Media, 2012.

\bibitem[Littman et~al.(2016)Littman, Wrubel, and Kerchner]{PDI7IN_2016}
Justin Littman, Laura Wrubel, and Daniel Kerchner.
\newblock 2016 united states presidential election tweet ids, 2016.
\newblock URL \url{https://doi.org/10.7910/DVN/PDI7IN}.

\bibitem[Marks and Zaman(2017)]{marks2017building}
Christopher Marks and Tauhid Zaman.
\newblock Building a location-based set of social media users.
\newblock \emph{arXiv preprint arXiv:1711.01481}, 2017.

\bibitem[Martins and Galam(2013)]{martins2013building}
Andr{\'e}~CR Martins and Serge Galam.
\newblock Building up of individual inflexibility in opinion dynamics.
\newblock \emph{Physical Review E}, 87\penalty0 (4):\penalty0 042807, 2013.

\bibitem[Messias et~al.(2013)Messias, Schmidt, Oliveira, and
  Benevenuto]{messias2013you}
Johnnatan Messias, Lucas Schmidt, Ricardo Oliveira, and Fabr{\'\i}cio
  Benevenuto.
\newblock You followed my bot! transforming robots into influential users in
  twitter.
\newblock \emph{First Monday}, 18\penalty0 (7), 2013.

\bibitem[Mobilia(2003)]{mobilia2003does}
Mauro Mobilia.
\newblock Does a single zealot affect an infinite group of voters?
\newblock \emph{Physical review letters}, 91\penalty0 (2):\penalty0 028701,
  2003.

\bibitem[Mobilia et~al.(2007)Mobilia, Petersen, and Redner]{mobilia2007role}
Mauro Mobilia, A~Petersen, and Sidney Redner.
\newblock On the role of zealotry in the voter model.
\newblock \emph{Journal of Statistical Mechanics: Theory and Experiment},
  2007\penalty0 (08):\penalty0 P08029, 2007.

\bibitem[M{\o}nsted et~al.(2017)M{\o}nsted, Sapie{\.z}y{\'n}ski, Ferrara, and
  Lehmann]{monsted2017evidence}
Bjarke M{\o}nsted, Piotr Sapie{\.z}y{\'n}ski, Emilio Ferrara, and Sune Lehmann.
\newblock Evidence of complex contagion of information in social media: An
  experiment using twitter bots.
\newblock \emph{PloS one}, 12\penalty0 (9):\penalty0 e0184148, 2017.

\bibitem[Moussa{\"\i}d et~al.(2013)Moussa{\"\i}d, K{\"a}mmer, Analytis, and
  Neth]{moussaid2013social}
Mehdi Moussa{\"\i}d, Juliane~E K{\"a}mmer, Pantelis~P Analytis, and
  Hansj{\"o}rg Neth.
\newblock Social influence and the collective dynamics of opinion formation.
\newblock \emph{PloS one}, 8\penalty0 (11):\penalty0 e78433, 2013.

\bibitem[Olshevsky and Tsitsiklis(2009)]{olshevsky2009convergence}
Alex Olshevsky and John~N Tsitsiklis.
\newblock Convergence speed in distributed consensus and averaging.
\newblock \emph{SIAM Journal on Control and Optimization}, 48\penalty0
  (1):\penalty0 33--55, 2009.

\bibitem[Paradise et~al.(2017)Paradise, Shabtai, Puzis, Elyashar, Elovici,
  Roshandel, and Peylo]{paradise2017creation}
Abigail Paradise, Asaf Shabtai, Rami Puzis, Aviad Elyashar, Yuval Elovici,
  Mehran Roshandel, and Christoph Peylo.
\newblock Creation and management of social network honeypots for detecting
  targeted cyber attacks.
\newblock \emph{IEEE Transactions on Computational Social Systems}, 4\penalty0
  (3):\penalty0 65--79, 2017.

\bibitem[Parlapiano and Lee(2018)]{ref:russianbots}
Alicia Parlapiano and C.~Lee, Jasmine.
\newblock {The Propaganda Tools Used by Russians to Influence the 2016
  Election}.
\newblock \emph{The New York Times}, February 2018.
\newblock URL
  \url{https://www.nytimes.com/interactive/2018/02/16/us/politics/russia-propaganda-election-2016.html}.

\bibitem[Price(2018)]{ref:russianbots_feinstein}
Molly Price.
\newblock {Democrats urge Facebook and Twitter to probe Russian bots }.
\newblock \emph{CNET}, January 2018.
\newblock URL
  \url{https://www.cnet.com/news/facebook-and-twitter-asked-again-to-investigate-russian-bots/}.

\bibitem[{Python}()]{wordsegment}
{Python}.
\newblock {Python Word Segmentation}.
\newblock \url{http://www.grantjenks.com/docs/wordsegment/}.
\newblock Accessed: 2018-08-14.

\bibitem[Ratkiewicz et~al.(2011)Ratkiewicz, Conover, Meiss, Gon{\c{c}}alves,
  Flammini, and Menczer]{ratkiewicz2011detecting}
Jacob Ratkiewicz, Michael Conover, Mark~R Meiss, Bruno Gon{\c{c}}alves,
  Alessandro Flammini, and Filippo Menczer.
\newblock Detecting and tracking political abuse in social media.
\newblock \emph{ICWSM}, 11:\penalty0 297--304, 2011.

\bibitem[Rogers and Bhowmik(1970)]{rogers1970homophily}
Everett~M Rogers and Dilip~K Bhowmik.
\newblock Homophily-heterophily: Relational concepts for communication
  research.
\newblock \emph{Public opinion quarterly}, 34\penalty0 (4):\penalty0 523--538,
  1970.

\bibitem[Rubin(2005)]{rubin2005causal}
Donald~B Rubin.
\newblock Causal inference using potential outcomes: Design, modeling,
  decisions.
\newblock \emph{Journal of the American Statistical Association}, 100\penalty0
  (469):\penalty0 322--331, 2005.

\bibitem[Shane(2017)]{shane2017fake}
S~Shane.
\newblock The fake americans russia created to influence the election.
\newblock \emph{The New York Times}, 7, 2017.

\bibitem[Sood and Redner(2005)]{sood2005voter}
Vishal Sood and Sidney Redner.
\newblock Voter model on heterogeneous graphs.
\newblock \emph{Physical review letters}, 94\penalty0 (17):\penalty0 178701,
  2005.

\bibitem[Stein et~al.(2011)Stein, Chen, and Mangla]{stein2011facebook}
Tao Stein, Erdong Chen, and Karan Mangla.
\newblock Facebook immune system.
\newblock In \emph{Proceedings of the 4th Workshop on Social Network Systems},
  page~8. ACM, 2011.

\bibitem[Summers(2017{\natexlab{a}})]{blm2016DataSet}
Ed~Summers.
\newblock https://archive.org/details/blacklivesmatter-tweets-2016.txt.
\newblock 2017{\natexlab{a}}.

\bibitem[Summers(2017{\natexlab{b}})]{macronLeaksDataSet}
Ed~Summers.
\newblock https://archive.org/details/macronleakstweets.
\newblock 2017{\natexlab{b}}.

\bibitem[Syeed(2017)]{bbmg17}
Nafeesa Syeed.
\newblock
  https://www.bloomberg.com/news/articles/2017-09-01/russia-linked-bots-hone-online-attack-plans-for-2018-u-s-vote.
\newblock \emph{Bloomberg}, 2017.

\bibitem[Thomas et~al.(2011)Thomas, Grier, Song, and
  Paxson]{thomas2011suspended}
Kurt Thomas, Chris Grier, Dawn Song, and Vern Paxson.
\newblock Suspended accounts in retrospect: an analysis of twitter spam.
\newblock In \emph{Proceedings of the 2011 ACM SIGCOMM conference on Internet
  measurement conference}, pages 243--258. ACM, 2011.

\bibitem[Timberg and Dwoskin(2018)]{ref:twitter_70mbots}
Craig Timberg and Elizabeth Dwoskin.
\newblock {Twitter is sweeping out fake accounts like never before, putting
  user growth at risk}.
\newblock \emph{The Washington Post}, July 2018.
\newblock URL
  \url{https://www.washingtonpost.com/technology/2018/07/06/twitter-is-sweeping-out-fake-accounts-like-never-before-putting-user-growth-risk/}.

\bibitem[Tran et~al.(2009)Tran, Min, Li, and Subramanian]{tran2009sybil}
Dinh~Nguyen Tran, Bonan Min, Jinyang Li, and Lakshminarayanan Subramanian.
\newblock Sybil-resilient online content voting.
\newblock In \emph{NSDI}, volume~9, pages 15--28, 2009.

\bibitem[Tsitsiklis et~al.(1986)Tsitsiklis, Bertsekas, and
  Athans]{tsitsiklis1986distributed}
John Tsitsiklis, Dimitri Bertsekas, and Michael Athans.
\newblock Distributed asynchronous deterministic and stochastic gradient
  optimization algorithms.
\newblock \emph{IEEE transactions on automatic control}, 31\penalty0
  (9):\penalty0 803--812, 1986.

\bibitem[Tsitsiklis(1984)]{tsitsiklis1984problems}
John~Nikolas Tsitsiklis.
\newblock Problems in decentralized decision making and computation.
\newblock Technical report, MASSACHUSETTS INST OF TECH CAMBRIDGE LAB FOR
  INFORMATION AND DECISION SYSTEMS, 1984.

\bibitem[Twitter(2012)]{ref:twitterSearchAPI}
Twitter.
\newblock Using the twitter search api.
\newblock https://dev.twitter.com/docs/using-search, October 2012.

\bibitem[Vassio et~al.(2014)Vassio, Fagnani, Frasca, and
  Ozdaglar]{vassio2014message}
Luca Vassio, Fabio Fagnani, Paolo Frasca, and Asuman Ozdaglar.
\newblock Message passing optimization of harmonic influence centrality.
\newblock \emph{IEEE transactions on control of network systems}, 1\penalty0
  (1):\penalty0 109--120, 2014.

\bibitem[Viswanath et~al.(2014)Viswanath, Bashir, Crovella, Guha, Gummadi,
  Krishnamurthy, and Mislove]{viswanath2014towards}
Bimal Viswanath, Muhammad~Ahmad Bashir, Mark Crovella, Saikat Guha, Krishna~P
  Gummadi, Balachander Krishnamurthy, and Alan Mislove.
\newblock Towards detecting anomalous user behavior in online social networks.
\newblock In \emph{USENIX Security Symposium}, pages 223--238, 2014.

\bibitem[Vosoughi et~al.(2018)Vosoughi, Roy, and Aral]{vosoughi2018spread}
Soroush Vosoughi, Deb Roy, and Sinan Aral.
\newblock The spread of true and false news online.
\newblock \emph{Science}, 359\penalty0 (6380):\penalty0 1146--1151, 2018.

\bibitem[Wang(2010)]{wang2010detecting}
Alex~Hai Wang.
\newblock Detecting spam bots in online social networking sites: a machine
  learning approach.
\newblock In \emph{IFIP Annual Conference on Data and Applications Security and
  Privacy}, pages 335--342. Springer, 2010.

\bibitem[Wang et~al.(2012)Wang, Mohanlal, Wilson, Wang, Metzger, Zheng, and
  Zhao]{wang2012social}
Gang Wang, Manish Mohanlal, Christo Wilson, Xiao Wang, Miriam Metzger, Haitao
  Zheng, and Ben~Y Zhao.
\newblock Social turing tests: Crowdsourcing sybil detection.
\newblock \emph{arXiv preprint arXiv:1205.3856}, 2012.

\bibitem[Wang et~al.(2013)Wang, Konolige, Wilson, Wang, Zheng, and
  Zhao]{wang2013you}
Gang Wang, Tristan Konolige, Christo Wilson, Xiao Wang, Haitao Zheng, and Ben~Y
  Zhao.
\newblock You are how you click: Clickstream analysis for sybil detection.
\newblock In \emph{USENIX Security Symposium}, volume~9, pages 1--008, 2013.

\bibitem[Wintour(2018)]{ref:brexit_russia}
Patrick Wintour.
\newblock Russian bid to influence brexit vote detailed in new us senate
  report.
\newblock \emph{The Guardian}, January 2018.
\newblock URL
  \url{https://www.theguardian.com/world/2018/jan/10/russian-influence-brexit-vote-detailed-us-senate-report}.

\bibitem[Wu and Huberman(2004)]{wu2004social}
Fang Wu and Bernardo~A Huberman.
\newblock Social structure and opinion formation.
\newblock \emph{arXiv preprint cond-mat/0407252}, 2004.

\bibitem[Yang et~al.(2014)Yang, Wilson, Wang, Gao, Zhao, and
  Dai]{yang2014uncovering}
Zhi Yang, Christo Wilson, Xiao Wang, Tingting Gao, Ben~Y Zhao, and Yafei Dai.
\newblock Uncovering social network sybils in the wild.
\newblock \emph{ACM Transactions on Knowledge Discovery from Data (TKDD)},
  8\penalty0 (1):\penalty0 2, 2014.

\bibitem[Yardi et~al.(2009)Yardi, Romero, Schoenebeck,
  et~al.]{yardi2009detecting}
Sarita Yardi, Daniel Romero, Grant Schoenebeck, et~al.
\newblock Detecting spam in a twitter network.
\newblock \emph{First Monday}, 15\penalty0 (1), 2009.

\bibitem[Yildiz et~al.(2013)Yildiz, Ozdaglar, Acemoglu, Saberi, and
  Scaglione]{yildiz2013binary}
Ercan Yildiz, Asuman Ozdaglar, Daron Acemoglu, Amin Saberi, and Anna Scaglione.
\newblock Binary opinion dynamics with stubborn agents.
\newblock \emph{ACM Transactions on Economics and Computation}, 1\penalty0
  (4):\penalty0 19, 2013.

\bibitem[Yu et~al.(2006)Yu, Kaminsky, Gibbons, and Flaxman]{yu2006sybilguard}
Haifeng Yu, Michael Kaminsky, Phillip~B Gibbons, and Abraham Flaxman.
\newblock Sybilguard: defending against sybil attacks via social networks.
\newblock In \emph{ACM SIGCOMM Computer Communication Review}, volume~36, pages
  267--278. ACM, 2006.

\bibitem[Yu et~al.(2008)Yu, Gibbons, Kaminsky, and Xiao]{yu2008sybillimit}
Haifeng Yu, Phillip~B Gibbons, Michael Kaminsky, and Feng Xiao.
\newblock Sybillimit: A near-optimal social network defense against sybil
  attacks.
\newblock In \emph{Security and Privacy, 2008. SP 2008. IEEE Symposium on},
  pages 3--17. IEEE, 2008.

\bibitem[Zabih and Kolmogorov(2004)]{zabih2004spatially}
Ramin Zabih and Vladimir Kolmogorov.
\newblock Spatially coherent clustering using graph cuts.
\newblock In \emph{Computer Vision and Pattern Recognition, 2004. CVPR 2004.
  Proceedings of the 2004 IEEE Computer Society Conference on}, volume~2, pages
  II--II. IEEE, 2004.

\bibitem[Zangerle and Specht(2014)]{zangerle2014sorry}
Eva Zangerle and G{\"u}nther Specht.
\newblock Sorry, i was hacked: a classification of compromised twitter
  accounts.
\newblock In \emph{Proceedings of the 29th Annual ACM Symposium on Applied
  Computing}, pages 587--593. ACM, 2014.

\end{thebibliography}
\bibliographystyle{plainnat}

\ECSwitch

\ECHead{Electronic Companion for Detecting Bots and Assessing Their Impact in Social Networks }

In this E-Companion we provide additional data analysis for the manuscript ``Detecting Bots and Assessing Their Impact in Social Networks''.

\section{Robustness of Ising Model Bot Detection Algorithm to Parameter Values}\label{sec:robust_ising}
In this section we present results showing the robustness of the Ising model bot detection algorithm to the variations in its parameter values.  We first look at the parameters associated with the link energies, and then the node energies.

\subsection{Link Energy Robustness} 
To show the robustness of the algorithm, we check how variations in the link energy parameter values affect the resulting AUC on the datasets.
The link energies are determined by the parameters $\gamma$, $\alpha_{in}$, $\alpha_{out}$, $\lambda_{00}$,  $\lambda_{01}$,  $\lambda_{10}$,  and $\lambda_{11}$.  We fix the parameters $\lambda_{10}=1$ and $\lambda_{10} = \lambda_{00}+\lambda_{11}-1$.  Therefore, we only analyze the robustness of the algorithm with respect to $\gamma$, $\alpha_{in}$, $\alpha_{out}$, $\lambda_{00}$, and  $\lambda_{11}$.  

To perform our robustness analysis, we vary one group of parameters while leaving all other parameters fixed.  The baseline parameter values are those presented in Section \ref{sec:roc_ising}: $(\lambda_{10},\lambda_{00},\lambda_{11},\lambda_{01}) = (0.44,0.61,0.83,1)$, $\gamma = 1$, and  $(\alpha_{out}, \alpha_{in})=(100,100)$ for all events except BLM 2016 where we set $(\alpha_{out},\alpha_{in})=(100,1000)$. We vary $\gamma$ over the values $\curly{0.1,1,10}$.   The degree parameters $(\alpha_{out},\alpha_{in})$ are varied over $\curly{50,100,200}\times\curly{50,100,200}$ except for BLM 2015, where we vary the parameters over
$\curly{50,100,200}\times\curly{500,1000,2000}$.  We vary  $(\lambda_{00},\lambda_{11})$  over
$\curly{0.6,0.7,0.8}\times\curly{0.6,0.7,0.8}$ with the heterophily/homophily constraint $\lambda_{00}\leq \lambda_{11}$.
We also check if the algorithm is robust to joint variations in the parameters.  To do this we randomly sample the parameters from the sets described rather than varying them one at a time.

The results of our robustness analysis are presented in Tables \ref{table:ising_robust_link_auc} and \ref{table:ising_random}.  For Table \ref{table:ising_robust_link_auc} each row shows the minimum, maximum, and mean values of the AUC as a group of parameters are varied.  As can be seen, the AUC is quite robust to the parameter values.  The $\gamma$ parameter has  no impact, which is expected because we use zero for the node energies, making the total energy proportional to $\gamma$.  Variations in $\alpha_{in},\alpha_{out},\lambda_{00},\lambda_{11}$ cause the AUC to change by only a few percent.     In Table \ref{table:ising_random} we show statistics of the AUC for 100 random draws of the algorithm parameters on the Pizzagate dataset.  As can be seen, even with joint parameter variation, the algorithm performance remains stable.

\begin{table}[h!]
	\begin{center}
		\caption{Mean, minimum, and maximum AUC of Ising model bot detection algorithm on the datasets as each group of parameters are varied.  Each row indicates the parameters that are varied.  The parameters that are not varied are set to  $(\lambda_{10},\lambda_{00},\lambda_{11},\lambda_{01}) = (0.44,0.61,0.83,1)$, $\gamma = 1$, and  $(\alpha_{out}, \alpha_{in})=(100,100)$ for all events except BLM 2016 where we set $(\alpha_{out},\alpha_{in})=(100,1000)$.  }\label{table:ising_robust_link_auc}
		\begin{tabular}{|l| c|c|c|c|}
			\hline
			Dataset & Varied parameter& Mean AUC&Minimum AUC &Maximum AUC \\ \hline
			
			\multirow{4}{*}{Pizzagate} & $\gamma$ & 0.91 & 0.91 & 0.90 \\\cline{2-5}
			 & $\alpha_{out},\alpha_{in}$ & 0.90 & 0.87 &0.91\\\cline{2-5} 			
			 & $\lambda_{00},\lambda_{11}$ & 0.89 & 0.83 & 0.91\\\specialrule{2.5pt}{1pt}{1pt}
			          
			\multirow{4}{*}{BLM 2015} & $\gamma$ & 0.67 & 0.67 & 0.67 \\\cline{2-5}
			& $\alpha_{out},\alpha_{in}$ & 0.69 & 0.67 & 0.72\\\cline{2-5} 			
			& $\lambda_{00},\lambda_{11}$ & 0.68 & 0.67 & 0.76\\\specialrule{2.5pt}{1pt}{1pt}
			
			\multirow{4}{*}{U.S. Presidential Debate} & $\gamma$ & 0.83 & 0.83 & 0.83 \\\cline{2-5}
			& $\alpha_{out},\alpha_{in}$ & 0.83 & 0.8 &0.85\\\cline{2-5} 			
			& $\lambda_{00},\lambda_{11}$ & 0.83 & 0.81 & 0.84\\\specialrule{2.5pt}{1pt}{1pt}	
			
			\multirow{4}{*}{Macron Leaks} & $\gamma$ & 0.84 & 0.84 & 0.84\\\cline{2-5}
			& $\alpha_{out},\alpha_{in}$ & 0.85 & 0.84 &0.88\\\cline{2-5} 			
			& $\lambda_{00},\lambda_{11}$ & 0.83 & 0.79 & 0.84\\\specialrule{2.5pt}{1pt}{1pt}
			
			\multirow{4}{*}{Hungary Elections} & $\gamma$ & 0.83 & 0.83 & 0.83 \\\cline{2-5}
			& $\alpha_{out},\alpha_{in}$ & 0.84 & 0.82 & 0.87 \\\cline{2-5} 			
			& $\lambda_{00},\lambda_{11}$ & 0.83 & 0.83 & 0.83\\\specialrule{2.5pt}{1pt}{1pt}
			
			\multirow{4}{*}{BLM 2016} & $\gamma$ & 0.91 & 0.91 & 0.91 \\\cline{2-5}
			& $\alpha_{out},\alpha_{in}$ & 0.90 & 0.88 & 0.91\\\cline{2-5} 			
			& $\lambda_{00},\lambda_{11}$ & 0.91 & 0.91 & 0.91\\\hline
	\end{tabular}
	\end{center}
\end{table}

\begin{table}[h!]
	\begin{center}
		\caption{Mean, minimum, and maximum AUC of Ising model bot detection algorithm on the Pizzagate dataset as the parameter values are jointly varied.  The AUC is calculated for 100 random samples of the parameters.}
		\label{table:ising_random}
		\begin{tabular}{|l|c|c|c|}
			\hline
			Dataset &  Mean AUC & Minimum AUC & Maximum AUC \\ \hline
			Pizzagate & 0.91       &  0.85               & 0.95\\ \hline
		\end{tabular}
	\end{center}
\end{table}


\subsection{Node Energy Robustness} \label{sec:node_energy_robust}
For each user or node in the retweet graph, we need a node energy value for each node label value.  Define a score $\pi_i\in[0,1]$ for a user $i$.  This can be viewed as the a priori probability that the user is a bot. The node energies are then set to $\phi(\mathbf x_i,0)=-\log(1-\pi_i)$ and $\phi(\mathbf x_i,1)=-\log(\pi_i)$.

We tested different choices for this score, which we list below.  
\begin{itemize}
	\item Zero: $\pi_i$ is 0.5 for all users. This makes the node energy independent of the label. Therefore, for simplicity we set the node energies to zero.
	
	\item Uniform: $\pi_i$ is drawn from a uniform distribution on $[0,1]$.
	
	\item BotOrNot: $\pi_i$ is set to the  probability the BotOrNot algorithm \citep{davis2016botornot} assigns to the user of being a bot.
\end{itemize}
The first two choices do not incorporate any information about the users and the retweet graph.  The third choice uses all the data needed to apply the BotOrNot algorithm.  In the case where the retweet graph has no edges, the Ising model algorithm would produce the same bot probabilities as BotOrNot with this choice for the node energies.

We show the resulting AUC scores on the datasets for these choices of node energies in Table \ref{table:ising_robust_node_auc}. For the uniform choice, we report the mean, minimum, and maximum AUC over ten random instances.   The link energy parameter values are those presented in Section \ref{sec:roc_ising}: $(\lambda_{10},\lambda_{00},\lambda_{11},\lambda_{01}) = (0.44,0.61,0.83,1)$, $\gamma = 1$, and  $(\alpha_{out}, \alpha_{in})=(100,100)$ for all events except BLM 2016 where we set $(\alpha_{out},\alpha_{in})=(100,1000)$.   One thing we see in the table is that the uniform choice does quite poorly many times.  This is most likely due to users who are not bots being given high values for $\pi_i$, or vice versa.  By comparison, the zero choice appears to perform much better.  The inclusion of the BotOrNot score provides a small improvement for datasets such as the Hungarian election, BLM 2015, and the U.S. presidential debate.  However, for BLM 2016 and Pizzagate the BotOrNot score actually reduces the AUC.  Therefore, we cannot conclude that inclusion of this prior information is always beneficial.  Overall, the zero versus BotOrNot choices for the node energies are very similar, with the exception of BLM 2015 and BLM 2016.  Given these findings, we would prefer the zero node energies because they require less data, are easier to compute, and seem to perform as well as more complex node energies.

\begin{table}[!hbt] \centering
	\caption{AUC measure of Ising model bot detection algorithm with different node energies applied to  Twitter datasets for different events.  For the uniform node energy choice, we report the mean, (minimum, maximum) AUC.}
	\label{table:ising_robust_node_auc}
	\centering
	\begin{tabular}{|l|c|c|c|}
		\hline
		Dataset      &AUC: zero&  AUC: uniform & AUC: BotOrNot\\\hline
		Pizzagate               & 0.91 & 0.83 (0.81, 0.86)  &0.90 \\\hline
		BLM 2015                & 0.67& 0.70 (0.68, 0.73)& 0.76\\\hline
		US presidential debate  &0.83& 0.58 (0.48, 0.63) & 0.84\\\hline
		Macron leaks            &0.84 & 0.83 (0.77, 0.85)& 0.85\\\hline
		Hungarian election      &0.83& 0.69 (0.60, 0.76) & 0.87\\\hline
		BLM 2016                &0.91& 0.62 (0.58, 0.68)  &0.67\\\hline
	\end{tabular}
\end{table}

\section{Keywords Used to Label Tweets for Neural Network}\label{sec:labeling_tweets}
The keywords used to label the neural network training data for the datasets are shown in Tables \ref{table:hashtag_choices_us}, \ref{table:hashtag_choices_BREXIT}, and \ref{table:hashtag_choices_YellowVests}.  The strings that do not correspond to words are unicode sequences for different images known as emojis.  For instance, the emoji in Table \ref{table:hashtag_choices_BREXIT} corresponds to the flag of the European Union.  

In Table \ref{table:hashtag_choices_YellowVests} there is a third column titled ``mixed''.  The words in this column were used to collect tweets to build the Gilets Jaunes dataset.  Recall from Section \ref{sec:data} that we included these words in our search query to obtain a larger number of tweets related to Gilets Jaunes.

\begin{table}[h!]
	\begin{center}
		\caption{Keywords used for the construction of the neural network training labels for the U.S. presidential debate dataset.  }
		\label{table:hashtag_choices_us}
		\begin{tabular}{|l|l|} 
			\hline
			\multicolumn{1}{|c|}{Pro-Trump}   & \multicolumn{1}{|c|}{Anti-Trump}  \\
			\hline
			presidenttrump, makeamericagreatagain, & StillWithHer, DemForce, \\
			killaryclinton, crookedhillary, & ImWithHer, ImStillWithHer,\\
			maga, donaldtrump2016, libtards, & TrumpRussia, TheRussiansHackedUs,\\		 
			 obamasucks, voterepublican, & HillaryIsMyPresident, AlwaysWithHer, \\
			 presidentdonaldtrump, votetrump,&demsinphilly, Demconvention,\\
			votedonald, 	votedonaldtrump,& BlackLivesMatter, BasketOfDeplorables, \\
			buildthatwall, draintheswamp, & BLM, NeverTrump,\\
			trumptrain, Trump2016, & DeleteYourAccount, TrumpTapes,\\
			Trump2020, tcot, BlueLivesMatter, & OHHillYes, Strongertogether,\\
			PresidentTrump, BuildTheWall, &hillary2016, hillarysupporter \\
			NeverHillary, TCOT, CCOT,  & factcheck, LastTimeTrumpPaidTaxes\\
			SethRich, PJNET, AllLivesMatter, & dnc, dems, dumptrump, \\
			
			POTUS, LockHerUp, & UniteBlue, ClintonKaine16, \\
			
			RedNationRising, IStandWithIsrael, FakeNews,  & hrc, Hillary, HillaryClinton, NotMyPresident \\
			BanIslam, ProIsrael, America1st, TeaParty,& \\
		    TrumpPence2020, DTS, 1a, 2a, prolife, & \\
			Benghazi, NRA, DemExit, Deplorable, BoycottNFL,  &  \\  
			TrumpPence16, TrumpPence, & \\
			TrumpPence2016, TRUMP2020& \\
			\hline
			
		\end{tabular}
	\end{center}
\end{table}

\begin{table}[h!]
	\begin{center}
		\caption{Keywords used for construction of the neural network training labels for the Brexit dataset.}
		\label{table:hashtag_choices_BREXIT}
		\begin{tabular}{|l|l|} 
			\hline
			\multicolumn{1}{|c|}{Pro-Brexit}   & \multicolumn{1}{|c|}{Anti-Brexit}  \\
			\hline
			BrexitmeansBrexit, SupportBrexit, & FBPE, StopBrexit, \\
			HardBrexit, Full Brexit, & strongerin, greenerin,\\
			LeaveMeansLeave, & intogether,\\
			Brexiter, Brexiteer, & infor, remain, \\ 
			antieu, Anti EU, & Bremain,  votein, \\
			no2eu,   wtobrexit, & incrowd, \\
			FullBrexitProperExit & yes2europe, \\
			ProBrexit, PlanAPlus, & exitfrombrexit,   Eunity, \\
			ChuckCheques, ChuckCheq,&  Forthemany,  DeeplyUnhelpful,\\
			voteleave, votedleave,&  WATON,  ABTV,\\
			ivotedleave, &  EUsupergirl,\\
			voteout,	votedout, &  FBSI, NHSLove, \\
			pro-brexit,	pro brexit, probrexit, &   U0001f1eaU0001f1fa (EU flag emoji),  \\
			takebackcontrol,	betteroffout, &\\
			StandUp4Brexit, WeAreLeaving &  \\
			\hline
			
		\end{tabular}
	\end{center}
\end{table}

\begin{table}[h!]
	\begin{center}
		\caption{Keywords used for the construction of the neural network training labels for the Gilets Jaunes dataset.  These keywords were also used to collect tweets to build the Gilets Jaunes dataset.}
		\label{table:hashtag_choices_YellowVests}
		\begin{tabular}{|l|l|l|} 
			\hline
			\multicolumn{1}{|c|}{Pro-Gilets Jaunes}   & \multicolumn{1}{|c|}{Anti-Gilets Jaunes}  & \multicolumn{1}{|c|}{Mixed} \\
			\hline
			YellowVests, & giletsbleu, & GrandD\'ebat, \\
			violencepoliciere, & cr\'etinsjaunes, & GrandDebat,\\
			\'EtatDeDroit, & cretinsjaunes, & EmmanuelMacron, \\
			EtatDeDroit,  & STOP\c caSuffit,& Macron\\
			r\'epression, & stopcasuffit, &\\
			\'etatpolicier, & TouchePasAMonEurope,&\\
			EtatPolicier, & CetteFoisJeVote,&\\
			Anti EU, & EnsembleavecMacron,&\\
			Acte16, Acte17, Acte18,  & SoutienAuPr\'esidentMacron,&\\
			MacronDemission, Frexit & SoutienAuPresidentMacron,&\\
			& U0001f1eaU0001f1fa (EU flag emoji),&\\
			& U0001F1EBU0001F1F7 (French flag emoji)\\
			\hline
			
		\end{tabular}
	\end{center}
\end{table}
\section{Details of Neural Network }\label{sec:neural_network}
To asses the opinion of a tweet, we used a convolutional neural network architecture. Each tweet is first preprocessed in two versions and sent to two channels in the neural network. The model architecture  was inspired by  \cite{kim2014convolutional}.  For convenience we again show this neural network architecture in  Figure \ref{fig:CNN_ec}.  Their approach was to train a text classification model on two different word embeddings of the same text: one static channel comprised of embeddings using word2vec \citep{goldberg2014word2vec} and another channel which is the output of an embedding layer.

\begin{figure}[h!]
	\centering
	\includegraphics[scale=1.2]{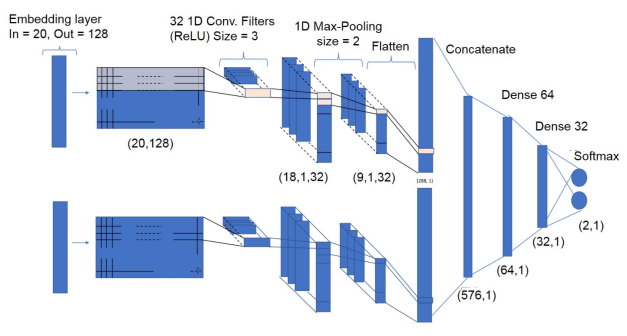}
	\caption{Diagram of the neural network architecture from \cite{kim2014convolutional} used to learn tweet opinions.  }
	\label{fig:CNN_ec}
\end{figure}

Each tweet is pre-processed into two one-hot encodings (see Section \ref{sec:preprocessing}).
Then, each version of the processed tweet goes through its own embedding layer \textit{(dimension dense embedding = 128)} that will then output two separate channels, each of size (20, 128). Each channel will go through its own separate 32 1D-convolution filters \textit{(kernel size = 3, stride = 1, padding = ‘valid’)}. Convolution filters enable one to represent n-grams  and learn shared parameters by convolving on various parts of the tweet. This prevents overfitting and enables one to learn translation invariant features. We then use a ReLU activation which is known to provide nice gradients for optimization and alleviate the problem of vanishing gradients. After the activation, we implement 1D max-pooling layers (pool size = 2). Pooling enables one to reduce computational cost, and enhance translational invariance by focusing on parts of the input where signals are the strongest. After pooling, we  use a flattening layer. The resulting output is two (288,1) layers that we concatenate to form a (576,1) layer. This layer then goes through two fully connected layers with a ReLU activation and 64 and 32 units, respectively. The final layer is a softmax layer that outputs the probability of the tweet's opinion being equal to one.


\subsection{Data Pre-processing for Neural Network}\label{sec:preprocessing}
Before being used to train the neural network, each tweet goes through a processing phase where we remove punctuation and stopwords and convert it into a format that the network can process. Each processed tweet is then converted  into two versions.
One version keeps hashtags as they are. This results in a one-hot encoding  vector of size $|\mathcal{V}|$, where $\mathcal{V}$ is the vocabulary of words when hashtags are left as they are.
The second encoding splits hashtags into actual words. This  results in a one-hot encoding of size $|\mathcal{V}^*|$, where $\mathcal{V}^*$ is the vocabulary of words when hashtags are broken down into separate words.

For example, \textit{I hope @candidate\_x will be our next president \#voteforcandidate\_x \#hatersgonnahate.} will be converted into two versions: 
\begin{itemize}
	\item[-] \textit{I hope candidate\_x will be our next president voteforcandidate\_x hatersgonnahate} 
	\item[-] \textit{I hope candidate\_x will be our next president vote for candidate\_x haters gonna hate}.
\end{itemize}
We  do this in order to prevent the neural network from being a lazy learner which only learns from the hashtags. 
This can also bring in more information since words are usually built on roots. For example, the commonly used hashtag  \emph{\#standUpForBrexit} will be broken down into stand + up + for + br + exit, hence  conveying the idea of exit as a good thing. If a new tweet is posted and mentions \emph{the necessary exit from the EU} then it will receive a score closer to pro-Brexit.

The hashtag splitting was done using the WordSegment library in Python \citep{wordsegment}. The sequence length of the tweets was set to 20 tokens (i.e. words). Any tweet with more than 20 tokens is truncated, while tweets with less than 20 tokens are padded with zeros.

\section{Robustness to Stubbornness Intervals}\label{sec:robustness_tests}
To make sure the network opinion shifts were robust to the choice of stubborn interval, we recalculated the opinions using several different intervals.  We did this for the U.S. presidential debate and Gilets Jaunes datasets because we saw large bot induced shifts here.  The resulting mean non-stubborn opinions for each choice of stubborn interval is shown in Tables \ref{table:thresholds_us} and \ref{table:thresholds_giletsjaunes}.

For the U.S. presidential debate dataset, we see that the mean opinion shifts are robust to the precise value of the stubbornness thresholds except for the $[0,0.15]$ and $[0.85,1]$ interval.  Here we see a smaller opinion shift.  At this threshold the number of stubborn users represent 16\% of the network, while in the other cases less than 12\% of the users are stubborn.  If the stubbornness intervals are too large, then the definition of stubborn comes into question.  Therefore, it is not useful to have too large stubbornness intervals. In fact we see in Table \ref{table:thresholds_us} that for narrower stubbornness intervals the opinion shift is not sensitive to the precise threshold values.

The results for the Gilets Jaunes dataset in Table \ref{table:thresholds_giletsjaunes} show a similar behavior.  For larger thresholds,  9\% to 14\% of the users are stubborn, versus 4\% for the $[0.0,0.1],[0.9,1.0]$ intervals.  We also see that the resulting shift decreases by 0.08. However, as with the U.S. presidential debate dataset, for these larger intervals the designated stubborn users may not in fact be stubborn.  

The conclusion of our robustness analysis is that as long as perturbations to the stubbornness intervals do not drastically change the number of of stubborn users, the opinions will also not change.  Therefore, the robustness of our results is tied to the shape of the user opinion distributions provided by the neural network.  Recall that we use the neural network opinions to identify stubborn users. Opinions with  cumulative distribution functions that are flat near the extreme values (opinions of zero and one) will be robust to the choice of stubbornness interval.

\begin{table}[h!]
	\begin{center}
		\caption{Mean opinions and number of stubborn users when all bots are removed using different stubborn thresholds in the U.S. presidential debate dataset.  There are 77,563 total users in the dataset. }\label{table:thresholds_us}
		\begin{tabular}{|c|c|c|c|c|c|} 
			\hline
			Lower   & Upper  & Mean    & Mean & Mean &Number of \\
			stubborn &  stubborn & opinion  & opinion&opinion&stubborn users\\
			interval &  interval & (no bots)  & (all bots)&shift &\\\hline
			$[0,0.075]$ & $[0.925]$ & 0.61 & 0.42 &0.19&5,284 \\\hline	
			$[0,0.09]$ & $[0.91,1]$ & 0.61 & 0.42&0.19&6,246 \\\hline
			$[0,0.10]$ & $[0.90,1]$ & 0.58 & 0.43&0.15 & 7,322\\  \hline
			$[0,0.11]$ & $[0.89,1]$ & 0.59 & 0.43 &0.16  & 8,113 \\  \hline
			$[0,0.125]$ & $[0.875,1]$ & 0.56 & 0.43 &0.13 & 9,493 \\  \hline
			$[0,0.15]$ & $[0.85,1]$ & 0.48 & 0.43 &0.05 & 12,145 \\  \hline
		\end{tabular}
	\end{center}
\end{table}

\begin{table}[h!]
	\begin{center}
		\caption{Mean opinions and number of stubborn users when all bots are removed using different stubborn thresholds in the Gilets Jaunes dataset.  There are 40,456 total users in the dataset. }\label{table:thresholds_giletsjaunes}
		\begin{tabular}{|c|c|c|c|c|c|} 
			\hline
			Lower   & Upper  & Mean    & Mean & Mean &Number of \\
			stubborn &  stubborn & opinion  & opinion&opinion&stubborn users\\
			interval &  interval & (no bots)  & (all bots)&shift &\\\hline
			$[0,0.10]$ & $[0.90,1]$ & 0.17 & 0.41 &0.24 & 1,801\\\hline
			$[0,0.15]$ & $[0.85,1]$ & 0.22 & 0.38 &0.16 & 3,505 \\\hline
			$[0,0.20]$ & $[0.80,1]$ & 0.28 & 0.38 &0.16 & 5,503 \\  \hline
		\end{tabular}
	\end{center}
\end{table}

\end{document}